\documentclass[fleqn,usenatbib]{mnras}

\usepackage{newtxtext,newtxmath}

\usepackage[T1]{fontenc}

\DeclareRobustCommand{\VAN}[3]{#2}
\let\VANthebibliography\thebibliography
\def\thebibliography{\DeclareRobustCommand{\VAN}[3]{##3}\VANthebibliography}

\usepackage{graphicx}	
\usepackage{amsmath}	
\usepackage{graphicx}	
\usepackage{amsmath}	
\usepackage{gensymb}
\usepackage{multirow}
\usepackage{subcaption}
\usepackage{ulem}
\usepackage{hyperref}

\title[Eccentric IMRIs in DM spikes]{Examining the effects of dark matter spikes on eccentric intermediate-mass ratio inspirals using \textit{N}-body simulations}

\author[Mukherjee et al.]{
Diptajyoti Mukherjee,$^{1}$\thanks{E-mail: diptajym@andrew.cmu.edu}
A.~Miguel~Holgado,$^1$
Go Ogiya$^3$, 
Hy Trac,$^{1,2}$ 
\\
$^{1}$McWilliams Center for Cosmology, Department of Physics, Carnegie Mellon University, Pittsburgh, PA 15213, USA\\
$^{2}$NSF AI Planning Institute for Physics of the Future, Carnegie Mellon University, Pittsburgh, PA 15213, USA \\
$^{3}$Institute for Astronomy, School of Physics, Zhejiang University, Hangzhou 310027, China
}

\date{Accepted XXX. Received YYY; in original form ZZZ}

\pubyear{2015}

\begin{document}
\label{firstpage}
\pagerange{\pageref{firstpage}--\pageref{lastpage}}
\maketitle

\begin{abstract}

Recent studies suggest that dark matter (DM) spikes around intermediate-mass black holes could cause observable dephasing in gravitational wave (GW) signals from Intermediate Mass Ratio Inspirals (IMRIs). Previous research primarily used non-self-consistent analytic methods to estimate the impact of DM spikes on eccentric IMRIs. Our study provides the first self-consistent treatment of this phenomenon using $N$-body simulations, incorporating Post-Newtonian effects up to the 2.5 order for accurate and robust results. Contrary to prior works, which posited that the cumulative effect of two-body encounters (dynamical friction; DF) is the primary mechanism for energy dissipation, we reveal that a three-body effect (slingshot mechanism) plays a more significant role in driving the binary system's energy loss and consequent orbital shrinkage. We find that binaries counter-rotating with respect to the DM spike merge faster, while co-rotating binaries merge slower, contrary to expectations from the DF theory. Using Fokker-Planck methods, we also assess the presence and detectability of spikes in realistic environments. When interacting with surrounding materials, DM spikes can have shallower slopes and lower densities than previously considered, leading to smaller signals and lower detection prospects via dephasing. Our results suggest that `deshifting' rather than dephasing might be a more optimistic signature, as it is more robust even in low-density environments.

\end{abstract}


\graphicspath{{./}{figures/}}



\section{Introduction}

The nature of dark matter (DM) remains one of the most pressing mysteries in modern astrophysics. While its existence has been inferred through a range of indirect observations such as galaxy rotational curves \citep[e.g.,][]{Rubin1970ApJ...159..379R,Rubin1978ApJ...225L.107R,Rubin1980ApJ...238..471R,Presic1996MNRAS.281...27P}, and gravitational lensing of galaxy clusters \citep[e.g.,][]{Tyson1990ApJ...349L...1T,Hammer1991ApJ...383...66H,Fort1992ApJ...399L.125F,LeFevre1994ApJ...422L...5L}, its properties and interactions remain largely unknown. Cosmological simulations suggest that DM resides in galactic halos and is distributed according to the Navarro-Frenk-White profile \citep[][see second reference for a review]{NFW1996ApJ...462..563N,Bertone2010pdmo.book.....B}. Galaxies, and therefore galactic halos, often contain massive black holes (MBHs) at the center. \citet{Gondolo1999PhRvL..83.1719G} proposed that the adiabatic growth of these MBHs would modify the dark matter profile near them and create extremely dense density spikes. The density profile of these DM spikes $\rho_{\mathrm{DM}}$ is parameterized as a power-law profile and is given as \citep[e.g.,][]{Gondolo1999PhRvL..83.1719G,Eda2013PhysRevLett.110.221101,Kavanagh2020PhRvD.102h3006K} 
\begin{gather} \label{eq:dens_profile}
    \rho_{\mathrm{DM}} (r) = \rho_{\mathrm{sp}} \left ( \frac{r_{\mathrm{sp}}}{r} \right )^{\gamma_{\mathrm{sp}}}
\end{gather}
where $r$ denotes the distance from the central MBH, $\rho_{\mathrm{sp}}$ is a normalization factor for the density profile, $\gamma_{\mathrm{sp}}$ characterizes the power-law profile of the spike and $r_{\mathrm{sp}}$ is a characteristic radius of the spike.Usually $\gamma_{\mathrm{sp}}$ is taken to be between 2 to 2.5. For a central MBH with a mass of $10^3 M_{\odot}$, under these parameters, $r_{\mathrm{sp}}\approx 0.5 \rm{pc}$ and the density $10^{-6}$ pc from the MBH is $\sim 10^{15} M_{\odot} \rm{pc}^{-3}$. 

Although the authors showed that such spikes would typically form around Supermassive Black Holes (SMBHs) with $M_{BH} \geq 10^6 M_{\odot}$, where $M_{BH}$ is the mass of the black hole (BH), later work \citep[e.g.,][]{Ullio2001PhRvD..64d3504U,Merritt2002PhRvL..88s1301M,Merritt2004PhRvL..92t1304M,Bertone2005PhRvD..72j3502B} found that spikes could be depleted via different astrophysical processes such as galaxy mergers, scattering with nearby stars, and off center seed BHs. While SMBHs are expected to undergo many mergers, the same cannot be said about Intermediate Mass Black Holes. DM spikes around Intermediate Mass Black Holes (IMBHs) with $10^3 \leq M_{BH} \leq 10^6$ are expected to survive as they undergo fewer mergers \citep[][]{ZhaoPhysRevLett.95.011301,BertonePhysRevD.72.103517}. Additionally, \citet{Ferrer2017PhRvD..96h3014F} showed that spinning IMBHs could form denser spikes. Thus, IMBHs are expected to be the predominant source for black holes surrounded by DM spikes.  

\begin{figure}
    \begin{center}
    \includegraphics[width=0.45\textwidth]{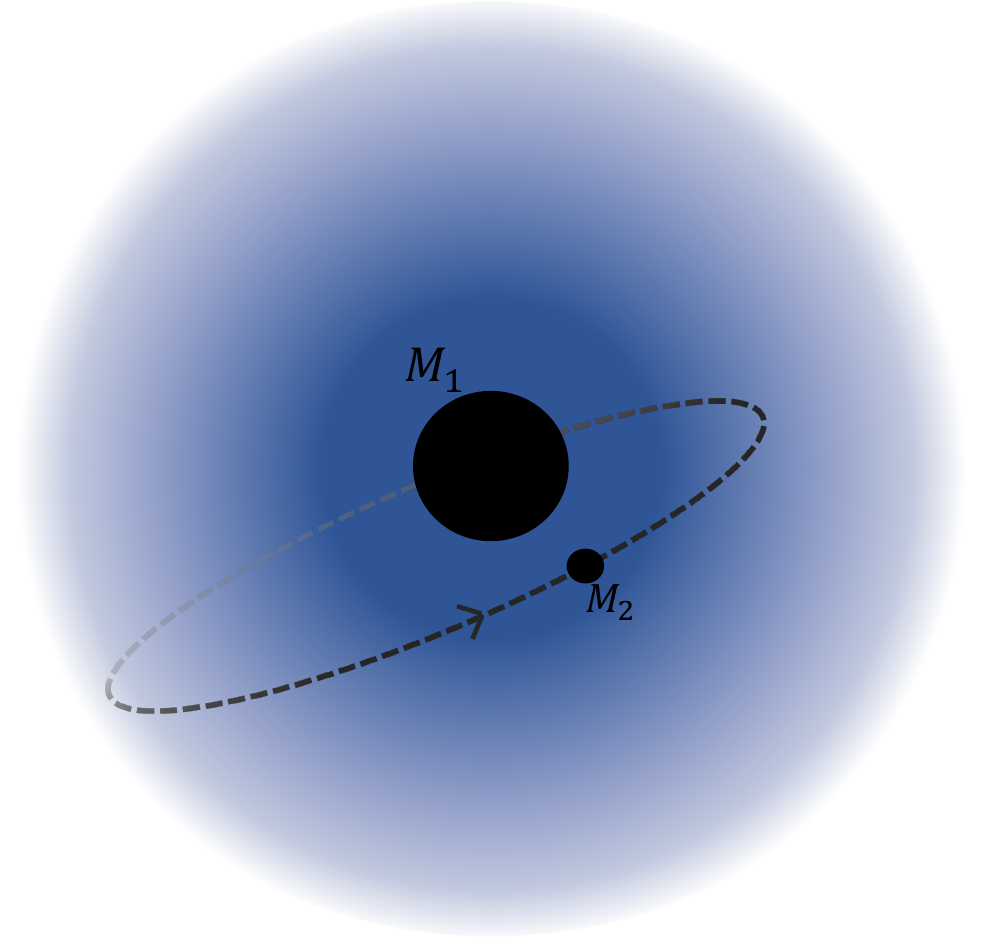}
    \caption{A visual representation of an eccentric IMRI embedded in a DM spike. The mass of the central BH is $M_1$ and that of the inspiraling object is $M_2$. This figure has been inspired by Figure 1 of \citet{Kavanagh2020PhRvD.102h3006K}. } 
    \label{fig:visual}
     \end{center}
\end{figure}

The advent of gravitational wave (GW) astronomy using LIGO-Virgo interferometers \citep[e.g.,][]{abbott2017gw170814,abbott2020gw190412,abbott2020gw190425,abbott2020gw190814} and pulsar timing array (PTA) \citep[e.g.,][]{Mingarelli2017NatAs...1..886M,Kelley2018MNRAS.477..964K,nanograv2023ApJ...951L...8A,nanograv2023ApJ...951L..50A} has opened up new avenues to detect and explore the properties of DM. In particular, it has been suggested that the effects of a DM spike could be imprinted on the GW signal of an Intermediate Mass Ratio Inspiral (IMRI) and has drawn a lot of interest \citep[e.g.,][]{Eda2013PhysRevLett.110.221101, Eda2015PhRvD..91d4045E,Yue2018PhRvD..97f4003Y,Yue2019ApJ...874...34Y,Kavanagh2020PhRvD.102h3006K,Becker2022PhRvD.105f3029B,Dai2022PhRvD.106f4003D}. A visual representation of such a system can be seen in Figure \ref{fig:visual}. IMRIs, formed from the inspiral of compact objects such as white dwarfs or neutron stars, or stellar mass black holes into IMBHs, will be detectable by future space based mHz GW detectors like LISA \citep{amaro2017laser} and TianQin \citep{luo2016tianqin} or deciHz GW detectors like DECIGO \citep{Decigo2019IJMPD..2845001K} or MAGIS \citep{Magis2021QS&T....6d4003A}.
GWs emanating from such systems are expected to be in the LISA and DECIGO  band for periods of months to years allowing environmental effects, including DM spikes, to play a major role in modifying the signal. Since matched filtering relies on a careful determination of the simulated signal to a few cycles over hundreds to thousands of cycles, it is imperative to take into account the environmental effects while numerically calculating the simulated GW signal \citep[e.g.,][]{Eda2013PhysRevLett.110.221101,Macedo2013ApJ...774...48M,Zwick2022MNRAS.511.6143Z,Coogan2022PhRvD.105d3009C,Baumann2022PhRvL.128v1102B,Zwick2023MNRAS.521.4645Z,Cole2023NatAs...7..943C,Cole2023PhRvD.107h3006C}

\cite{Eda2013PhysRevLett.110.221101} and \citet{Eda2015PhRvD..91d4045E} first proposed that the gravitational effects of the spike could leave an imprint on the GW signal of an IMRI. The spike particles would get scattered by the inspiraling object leading to an extra drag force experienced by the inspiraling object. This drag force has been attributed to dynamical friction \citep[DF;][]{Chandra1943ApJ....97..255C} on the inspiraling object. Although the DM mass contained in the spike is quite small relative to the mass of the central IMBH and the inspiraling object, it has been shown that the gravitational drag can lead to a phase shift in the GW signal over thousands of cycles.  Recent studies have shown that dephasing due to DF can lead to anywhere between $10^3-10^7$ fewer cycles which would be above the signal to noise detector theshold for LISA and could be detected and distinguished as an imprint of the DM spike surrounding the IMBH \citep[e.g.,][]{Eda2015PhRvD..91d4045E,Kavanagh2020PhRvD.102h3006K,Becker2022PhRvD.105f3029B,Dai2022PhRvD.106f4003D}. Different models of DM can lead to different spike parameters and as such change the amount of dephasing itself, which can be detected and distinguished. As such, GW astronomy provides a unique pathway to ascertain the presence and nature of DM itself. Detection of even one DM spike modified signal would be effective towards eliminating other theories of gravity such as modified Newtonian dynamics and can help constrain the particle nature of DM \citep[][]{Hannuksela2020PhRvD.102j3022H}. 

A careful determination of the the expected dephasing, therefore, is vital. While initial studies \citep[e.g.,][]{Eda2013PhysRevLett.110.221101,Eda2015PhRvD..91d4045E} used a static DM background, \citet{Kavanagh2020PhRvD.102h3006K} showed that the inspiral of the IMRI can inject energy into the spike itself, often comparable to the binding energy of the spike. As such the feedback of the IMRI onto the spike cannot be ignored. Through a semi-analytic framework called {\tt\string HaloFeedback}, \citet{Kavanagh2020PhRvD.102h3006K} showed that the including the back reaction from the IMRI can lead upto 100$\times$ reduction in expected amount of dephasing. Therefore, it is necessary to self-consistently follow the dynamics of the binary in order to calculate the amount of dephasing. While {\tt\string HaloFeedback} relies on the assumption that the binary is on a circular orbit at all times, studies \citep[e.g.,][]{Yue2018PhRvD..97f4003Y,Yue2019ApJ...874...34Y,Cardoso2021PhRvD.103b3015C,Dai2022PhRvD.106f4003D} showed that such an assumption might not hold in reality. A self-consistent framework has only been developed for binaries on circular orbits and studies including eccentric binaries neglect the backreaction from the binary to the spike \citep[e.g.,][]{Becker2022PhRvD.105f3029B,Dai2022PhRvD.106f4003D}. The feedback can not only affect the evolution of the semi-major axis but also the eccentricity which has a major impact on inspiral times due to GW emission. The rate of circularization of the IMRI has also recently been suggested as a signature of the spike \citep{Becker2022PhRvD.105f3029B} which would require a self-consistent framework that takes into account the effects of the feedback from the binary onto the spike and vice versa to be determined accurately.

Additionally, recent work has shown that the analytic Chandrasekhar approximation may lead to inconsistent evolution of semi-major axis and eccentricity due to the lack of inclusion of drag force from fast moving particles, which can be significant in certain cases \citep[][]{Fani2023arXiv230517281D}. As it is difficult to create a self-consistent analytic/semi-analytic framework to include these effects, we have to resort to $N$-body simulations.

$N$-body simulations, although expensive, are considered to be the gold standard of dynamical modelling. Previous approaches have rarely relied on $N$-body simulations as they are extremely computationally expensive. \citet{Kavanagh2020PhRvD.102h3006K} report that a simulation of 150 orbits of a 1 $M_{\odot}$ - 100 $M_{\odot}$ binary takes about 3 days using the $N$-body code {\tt\string GADGET-2} \citep[][]{Springel2005MNRAS.364.1105S}. This is very computationally expensive to study long term effects of the IMRI on the spike over tens of thousands of orbits. Therefore, in order to study the secular effects of the IMRI on the spike, different strategies are required. 

In our study we describe a novel $N$-body code that is over $100$ times faster than traditional $N$-body codes for simulating IMRIs embedded in DM spikes. Using our code, we present self-consistent results from eccentric IMRIs embdedded in DM spikes. 

In addition, we also study the effect of rotation in the DM spike which has not been done before. We argue that the inclusion of rotation is important since conservation of angular momentum would dictate that DM spikes should rotate upon formation from rotating galactic halos. Spinning IMBHs can also transfer angular momentum to the surrounding spike, making it rotate \citep{Ferrer2017PhRvD..96h3014F}.  
We systematically study the effects of the different post-Newtonian (PN) terms including precession and radiation reaction terms upto $2.5PN$.

We begin by describing our computational methods in Section \ref{sec:methods}, followed by the models of the IMRIs embedded in DM spikes in Section \ref{sec:models}. The results are then described in Section \ref{sec:results}, followed by discussions of our results and assumptions and conclusions in Sections \ref{sec:discussion}  and \ref{sec:conclusion} respectively.

\section{Computational Methods} \label{sec:methods}
The numerical evolution of the IMRI using $N$-body methods is an extremely computationally challenging problem owing to the enormous density of the spike near the central IMBH and the $\mathcal{O}(N^2)$ pairwise force calculations. An accurate evolution requires fine tuned time-steps for the particles near the central IMBH (primary) and inspiraling object (secondary) to resolve the scattering effects accurately. Previous studies have all relied on analytic/semi-analytic methods for long term simulations of the inspiraling IMRI. To the best of our knowledge, only \citet{Kavanagh2020PhRvD.102h3006K} performed $N$-body simulations of the binary embedded in a spike. However, the simulations were stopped after a period of $\sim 100$ orbits.  To understand realistic effects of the spike on the IMRI (and vice-versa) and to calibrate semi-analytic methods, we need long term $N$-body simulations. Here we present a novel $N$-body code \footnote{\url{https://github.com/dipto4/falcon\_dm/}} that is specifically tuned for simulations of IMRIs in DM spikes. 

\begin{figure}
    \begin{center}
    \includegraphics[width=0.45\textwidth]{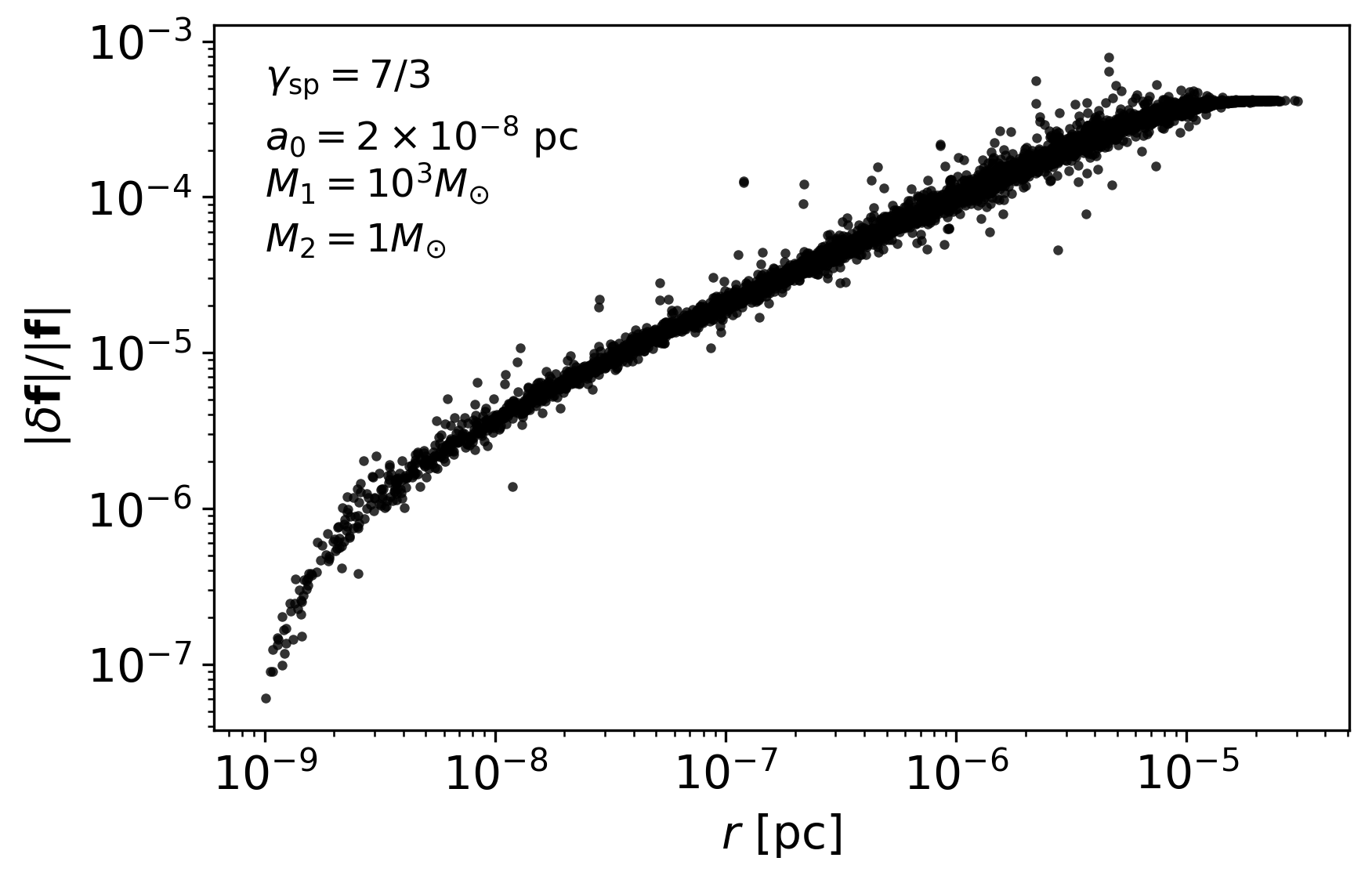}
    \caption{The relative force accuracy $|\delta \mathbf{f}|/|\mathbf{f}|$ of DM particles as a function of their separation $r$ from the central IMBH with mass $M_1=10^3 M_{\odot}$. The spike follows a $\gamma_{\mathrm{sp}}=7/3$ density profile. The inspiraling object, whose mass is $M_2=1 M_{\odot}$, is situated at an initial semi-major axis of $a_0=2\times10^{-8}$ pc. We find that in the region of interest, the force accuracy is $\leq 10^{-5}$, which is comparable to the the force accuracy obtained in tree/FMM based codes. } 
    \label{fig:accuracy}
     \end{center}
\end{figure}

A close analysis of the $N$-body simulation reveals that the bulk of computational time is spent calculating the forces of DM particles near the primary. Since the mass of DM in the region of interest is very small compared to the mass of the primary and the secondary, the self-gravity of the spike can be neglected. This allows us to safely neglect the interactions between the DM particles themselves. Full force calculations are only required for the primary and the secondary. Neglecting the self interaction of the spike effectively reduces the number of force calculations to $\mathcal{O}(N)$, speeding up the simulations massively. The mean relative force accuracy between a full $\mathcal{O}(N^2)$ calculation and our approximate method is $ \leq 10^{-5}$ in the region of interest as shown in Figure \ref{fig:accuracy}, on par with the force accuracies obtained by current Barnes-Hut tree \citep[][]{barnes1986hierarchical}  or Fast Multipole Method (FMM) \citep[][]{greengard1987fast,cheng1999fast,dehnen2002hierarchical,Qirong2021NewA...8501481Z}  based $N$-body codes. To verify the accuracy of our method, we compare the evolution of a $100 M_{\odot}-1 M_{\odot}$ binary in a DM spike with the FMM based code {\tt\string Taichi} \citep{Qirong2021NewA...8501481Z,Mukherjee2021ApJ...916....9M,Mukherjee2023MNRAS.518.4801M} and this method and find negligible differences after $\sim 1500$ orbits.  

To evolve the particles, we use a 2nd order hierarchical Hamiltonian splitting based integration scheme {\tt\string HOLD} in a Drift-Kick-Drift (DKD) fashion with symmetrized timesteps \citep{Pelupessy2012}. The usage of symmetrized timesteps ensures that there is no secular drift in energy and renders better energy conservation than other timestepping schemes \citep[e.g.,][]{Makino2006,Pelupessy2012,Mukherjee2021ApJ...916....9M}. For more information on the integration scheme, we refer the reader to \citet{Pelupessy2012}. The timesteps are controlled by a timestepping parameter $\eta$ which is proportional to the analytically calculated timestep. In our simulations, we set $\eta=0.025$. This results in a relative energy error of $\leq 10^{-10}$ per orbit, sufficient to follow the dynamics over few hundreds of thousands orbits. 

To account for relativistic effects, we add PN terms to the equations of motions for the primary and secondary. The PN equations of motion can be added to the standard Newtonian equation of motion as follows:
\begin{gather} \label{eq:pn}
    \mathbf{a} = \mathbf{a}_{\mathrm{Newtonian}} + \frac{1}{c^2} \mathbf{a}_{\mathrm 1PN} + \frac{1}{c^4} \mathbf{a}_{\mathrm 2PN} + \frac{1}{c^5} \mathbf{a}_{\mathrm 2.5PN} + ...
\end{gather}
where $\mathbf{a}_{nPN}$ represents the $n^{\mathrm{th}}$ PN term. We include up to $2.5PN$ terms in our calculations following equation 203 from \citet{Blanchet2014LRR....17....2B}. While the $1PN$ and $2PN$ terms only lead to precession of the binary's orbit, the $2.5PN$ term leads to GW radiation resulting in shrinkage of the orbit. The DM particles experience the Newtonian potential of the primary and the secondary. In principle this is not fully self-consistent since the DM particles will also experience PN effects in interactions with other DM particles. One can use the Einstein-Infield-Hoffman equations \citep[e.g.,][]{Will2014PhRvD..89d4043W,PortegiesZwart2022A&A...659A..86P} to self-consistently simulate the PN evolution of all the particles but this is beyond the scope of this study. 

A problem arises when adding velocity dependent forces like PN precession and radiation reaction terms since a leapfrog-like explicit splitting of the Hamiltonian is not achievable with velocity dependent forces. In such a scenario, the simple  DKD integration scheme cannot be utilized. An implicit method suggested by \citet{Mikkola2002CeMDA..84..343M} is popular but is quite inefficient. The implicit scheme requires multiple iterations to solve for the force. For highly non-linear vector fields, it proves to be quite computationally expensive. \citet{Hell2010CeMDA.106..143H} show that using a clever mathematical trick one can create an explicit DKD-like scheme by extending the phase space of variables by introducing an auxiliary velocity $\mathbf{w}$. An updated integration between step $t$ and $t+1$ using the auxiliary velocity $\mathbf{w}$ can now be written as:

\begin{gather}
    \mathbf{x}_{t+1/2} = \mathbf{x}_t + \frac{h}{2} \mathbf{v}_t \\
    \mathbf{w}_{t+1/2} = \mathbf{w}_t + \frac{h}{2} \mathbf{f}\left ( \mathbf{x}_{t+1/2}, \mathbf{v}_t \right) \\
    \mathbf{v}_{t+1} = \mathbf{v}_t + h \mathbf{f} \left ( \mathbf{x}_{t+1/2}, \mathbf{w}_{t+1/2}  \right ) \\
    \mathbf{w}_{t+1} = \mathbf{w}_{t+1/2} + \frac{h}{2} \mathbf{f} \left ( \mathbf{x}_{t+1/2}, \mathbf{v}_{t+1} \right) \\
    \mathbf{x}_{t+1} = \mathbf{x}_{t+1/2} + \frac{h}{2} \mathbf{v}_{t+1}
\end{gather}
where $\mathbf{x}$ is the position, $\mathbf{v}$ is the velocity, $\mathbf{f}$ is the force and $h$ is the timestep. We note that $\mathbf{w}_0 = \mathbf{v}_0$.

To test the integration scheme, we simulate the evolution of a $1000 M_{\odot} - 1 M_{\odot}$ binary in vacuum and compare the evolution of the eccentricity and semi-major axis to those derived using Peters' equations \citep{Peter1964PhRv..136.1224P} and find that the results are consistent with one another. We also compare the evolution of the binary against that from the publicly available regularization code {\tt\string SpaceHub} \citep{Wang2021MNRAS.505.1053W} which also includes PN terms upto 2.5PN and find no differences.

For three-body simulations, we utilize the 15th order Gauss-Radau integrator {\tt\string IAS15} \citep[][]{Rein2015MNRAS.446.1424R} from the {\tt\string rebound}  package \citep{rebound}. {\tt\string IAS15} can handle close encounters and is extremely accurate allowing for a closer and in-depth analysis into the dynamics of the scattering process of the binary with a DM particle. The energy is conserved to machine precision.

All simulations are performed on the Vera computing cluster utilizing {\tt\string AMD Epyc 7742} nodes. The majority of the full $N$-body simulations take about 100-120 core hours to finish to completion using a single core. The interactions between the secondary and DM particles are set to have zero softening while we set a relatively conservative softening value of $10^{-10}$ pc between the primary and DM particles in our simulations. This is equal to the Schwartzchild radius for the $10^3 M_{\odot}$ IMBH and one-tenth that for the $10^4 M_{\odot}$ IMBH. For the three-body simulations, all interactions are unsoftened. A set of simulations were run with softening between the DM particles and the secondary and no major differences were noticed between the simulations that used softening and those that did not. 
A small value of softening was necessary to prevent some particles near the primary from taking extremely small timesteps. The interaction between the primary and the secondary object is not softened.

\section{Models} \label{sec:models}

\begin{figure*}
    \begin{center}
    \includegraphics[width=1.0\textwidth]{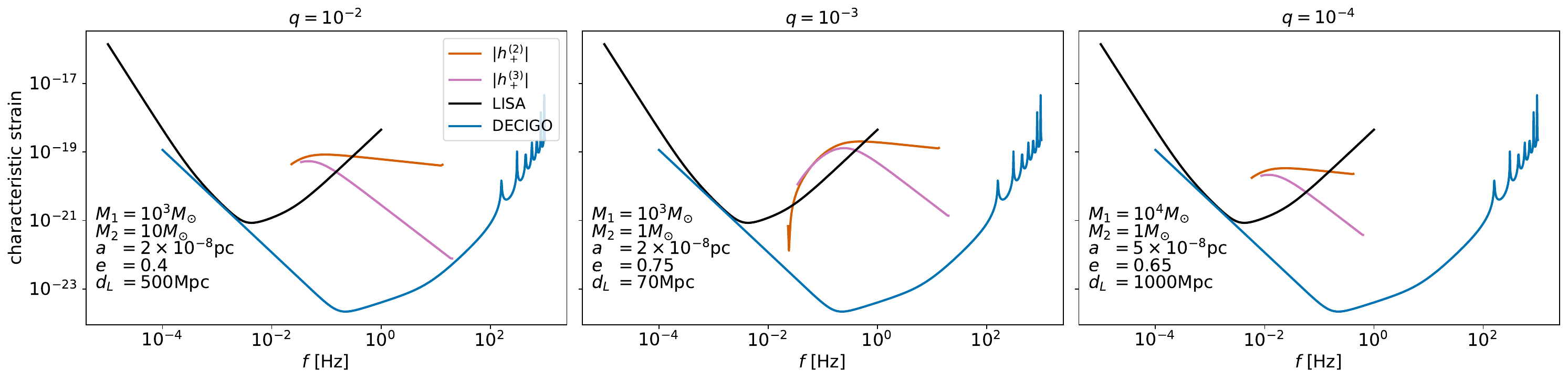}
    \caption{The characteristic strain of the IMRIs used as our initial conditions as a function of the frequency $f$ in vacuum for different mass-ratio $q$ models. The central IMBH has a mass of $M_1$ while the inspiraling object has a mass of $M_2$. The binary is defined by its initial semi-major axis $a$ and eccentricity $e$. The luminosity distance is denoted as $d_L$. All of the IMRIs presented here have a merger time of $\sim 5$ years. Since eccentric binaries radiate in multiple harmonics, we have plotted the strain from the second and third harmonics. In all of the scenarios, we find that the second harmonic has an equivalent or higher strain than the third harmonic. Even higher harmonics have lower strains and are harder to detect using LISA but would be detectable using DECIGO. }
    \label{fig:strain_ic}
     \end{center}
\end{figure*}

We are interested in scenarios wherein the IMRI is visible in the LISA/DECIGO band for a duration of $\sim$ 5 years. To understand the effects of varying eccentricity, primary and secondary masses, and density, we generate physically motivated models corresponding to IMRIs in vacuum where the GW signal is in the LISA/DECIGO band for $\sim 5$ years and above the LISA/DECIGO signal to noise threshold. We define the mass-ratio of the secondary to the primary $q$ as $q \equiv \frac{M_{2}}{M_{1}}$ where $M_{1}$ is the mass of the primary and $M_2$ is the mass of the secondary. We choose $M_1=10^3 M_{\odot} , 10^4 M_{\odot}$ and $M_2=1 M_{\odot}, 10 M_{\odot}$, representing three different scenarios with mass ratios $q=10^{-2},10^{-3},10^{-4}$ as depicted in Figure \ref{fig:strain_ic}. $M_2 = 1 M_{\odot}$ is representative of a scenario the inspiraling object is similar to a white dwarf of a neutron star whereas $M_2 = 10 M_{\odot}$ represents a scenario where the inspiraling object is a stellar mass black hole. The initial properties of the binary are characterized by its initial semi-major axis $a_0$ and eccentricity $e_0$. We set $a_0=2\times10^{-8}$ pc for $q=10^{-2},10^{-3}$ models and $a_0=5\times10^{-8}$ for the $q = 10^{-4}$ model. We also set $e_0=0.4$ for $q=10^{-2}$ model, $e_0=0.75$ for $q=10^{-3}$ model, and $e_0=0.65$ for $q=10^{-4}$ model. Figure \ref{fig:strain_ic} provides a representation of the strain versus frequency curves of the models superimposed over the strain-frequency curves for LISA and DECIGO.

\subsection{Non-rotating models}
\label{subsec:non_rot_models}

\begin{table*}
\centering
\resizebox{1.0\textwidth}{!}{%
\begin{tabular}{lllllllll}
\hline
$M_1$ [$M_{\odot}$] & $M_2$ [$M_{\odot}$]  & $q$  & $a_0$ [pc], $e_0$ & $\gamma_{\mathrm{sp}}, r_{\mathrm{sp}} [\rm{pc}]$ & $m_{\mathrm{DM}}$ [$M_{\odot}$]  & PN terms used & $t_{\mathrm{final}}$ & $N_{\mathrm{sims}}$ (for each config) \\
\hline
$10^3$  & $10.0$    & $10^{-2}$ & $2\times10^{-8}, 0.4$ & $7/3, 0.54$                   & $5 \times 10^{-5}$                & 2.5PN & merger   & 5                         \\
$10^3$ & $1.0$    & $10^{-3}$ & $2\times10^{-8}, 0.75$ & $7/3, 0.54$                   & $5 \times 10^{-5}$               & 2.5PN, all terms up to 2.5PN              & merger   & 5                         \\
$10^{3}$ & $1.0$    & $10^{-3}$ & $2\times10^{-8}, 0.75$  & $9/4, 0.54$                   & $10^{-5}$                                        & 2.5PN              & merger   & 5                         \\
$10^{3}$ & $1.0$    & $10^{-3}$ & $2\times10^{-8}, 0.75$  & $3/2, 0.46$                   & $10^{-11}$                                        & 2.5PN              & merger   & 5                         \\
$10^{4}$ & $1.0$    & $10^{-4}$ & $5\times10^{-8}, 0.65$  & $3/2, 0.98$                   & $2\times10^{-10}$                                        & 2.5PN              & merger   & 5                         \\

$10^{4}$ & $1.0$    & $10^{-4}$ & $5\times10^{-8}, 0.65$  & $7/3, 0.54$                   & $5 \times 10^{-5}$                & 2.5PN              & merger   & 5  \\
$10^{4}$ & $1.0$    & $10^{-4}$ & $5\times10^{-8}, 0.65$  & $7/3, 1.17$                   & $5 \times 10^{-5}$                & 2.5PN              & 2.5 yr   & 5  \\
\hline
\end{tabular}%
}
\caption{A summary of the initial conditions for our non-rotating models. The rotating models have similar initial conditions but are either in a net prograde motion or retrograde motion with respect to the motion of the IMRI. For more information on how to generate the initial conditions, we refer the reader to section \ref{sec:models}. }
\label{tab:summary_ic}
\end{table*}

\begin{figure}
    \begin{center}
    \includegraphics[width=0.45\textwidth]{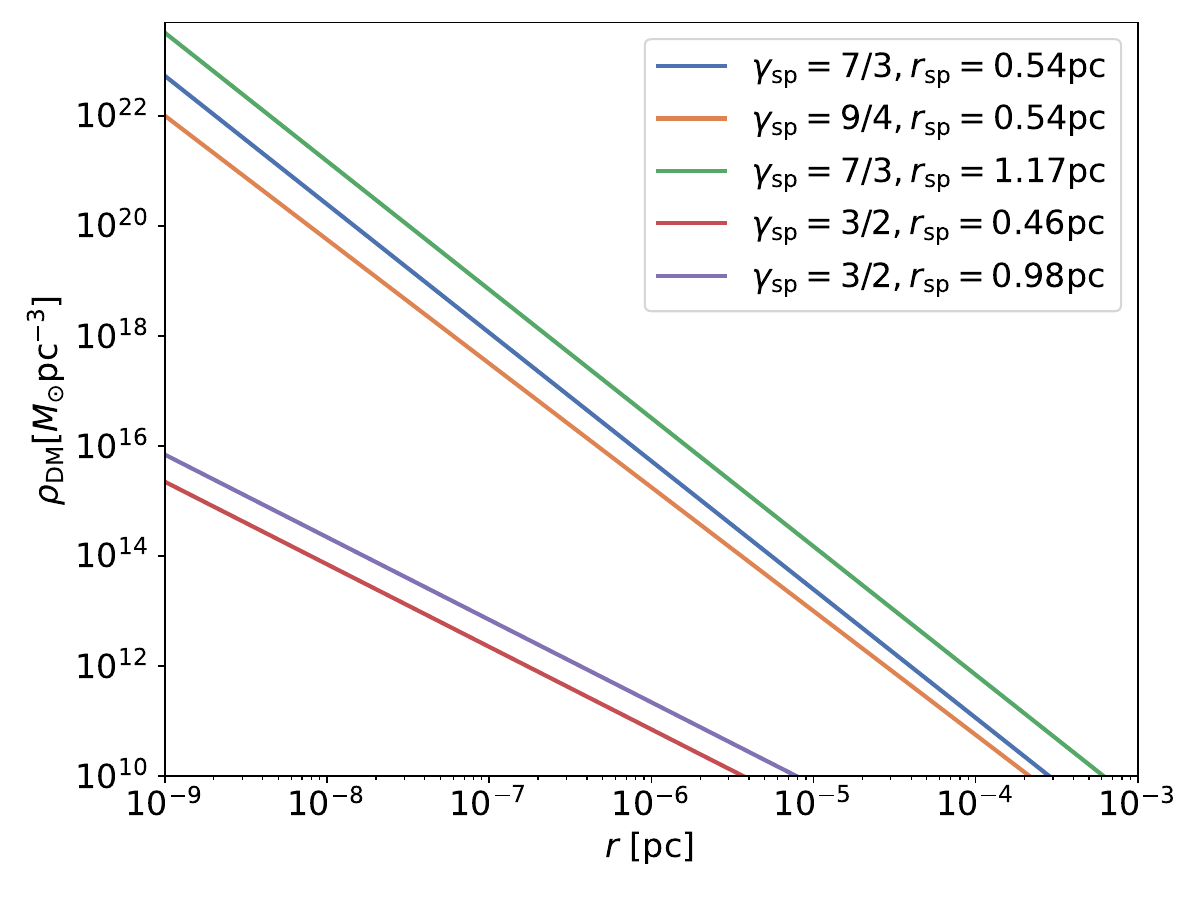}
    \caption{The density of the DM spike $\rho_{\mathrm{DM}}$ as a function of the distance from the central IMBH $r$. We use these density profiles to generate our $N$-body initial conditions.} 
    \label{fig:initial_density_profile}
     \end{center}
\end{figure}

To generate the $N$-body realizations of the DM spike, we utilize the galactic modeling toolkit {\tt\string Agama} \citep{Vasilev2019MNRAS.482.1525V}. Following previous studies \citep[e.g.,][]{Eda2015PhRvD..91d4045E,Kavanagh2020PhRvD.102h3006K}, we use equation \ref{eq:dens_profile} as our density profile. 
This profile is valid for all
for all radii $r > r_{\mathrm{ISCO}}$ where $r_{\mathrm{ISCO}}$ is the innermost stable circular orbit (ISCO) for the central IMBH. The density profile is taken to be 0 where $r < r_{\mathrm{ISCO}}$. Furthermore, $r_{\mathrm{sp}}$ is not considered to be a free parameter but is, instead, calculated as 
\begin{gather} \label{eq:rsp}
    r_{\mathrm{sp}} \approx \left ( \frac{(3-\gamma_{\mathrm{sp}}) 0.2^{3-\gamma_{\mathrm{sp}}} M_1}{2\pi \rho_{\mathrm{sp}}} \right)^{1/3}.
\end{gather}
For all our simulations we set $\rho_{\mathrm{sp}} = 226 M_{\odot} / \mathrm{pc}^3$ following \citet{Kavanagh2020PhRvD.102h3006K}. Using equation \ref{eq:rsp}, we find $r_{\mathrm{sp}} \approx 0.54$ pc when $M_1=10^3 M_{\odot}$ and $r_{\mathrm{sp}} \approx 1.17$ pc when $M_1=10^4 M_{\odot}$. Additionally, for most of our simulations we set $\gamma_{\mathrm{sp}}=7/3$. Such a spike profile can arise out of adiabatic growth of an IMBH in an NFW halo \citep{Eda2015PhRvD..91d4045E}. However, this is not universal and $\gamma_{\mathrm{sp}}$ depends on how the DM spike originates. For example, primordial BHs have been shown to produce a $\gamma_{\mathrm{sp}}=9/4$ spike around them \citep{Boudaud2021JCAP...08..053B}. 
To understand how the dephasing changes as $\gamma_{\mathrm{sp}}$ changes, we run a set of simulations with $\gamma_{\mathrm{sp}}=9/4$. Furthermore, to understand how the dephasing changes at a fixed density profile for different $q$, we perform a set of simulations with $M_1=10^4 M_{\odot}, r_{\mathrm{sp}}=0.54 \mathrm{pc}$ and $\gamma_{\mathrm{sp}}=7/3$. However, it is hard to robustly predict the existence of such dense spikes. In realistic environments where effects of stars cannot be neglected, DM ``crests'' are more likely. Such ``crests'' are formed due to the scattering of DM particles by the more massive more massive stellar particles and have been shown to produce a $\gamma_{\rm sp}=1.5$ profile \citep{Merritt2007PhRvD..75d3517M}. To understand if such profiles produce any observable dephasing effects, we perform a set of simulations with $\gamma_{\rm sp}=1.5$,  $M_1 = 10^3, 10^4 M_{\odot}$ and $M_2 = 1 M_{\odot}$. The density spikes considered in our simulations have been plotted in Figure \ref{fig:initial_density_profile}.
To generate models with finite total mass, we use an exponential truncation function with a truncation radius of $10^{-5}$ pc for $q=10^{-2},10^{-3}$ models and $10^{-6}$ pc for the $q=10^{-4}$ model. 
We verify that our choice of truncation radius does not affect the mass profile of the spike in the region of interest. We also compare our initial  $N$-body profiles to those from \citet{Kavanagh2020PhRvD.102h3006K} and find that the density profiles in the region of interest and the velocity profiles match. Additionally, we ensure that the density profile remains stable for the duration of the simulations. 

The mass of the DM particle, $m_{\mathrm{DM}}$, is chosen carefully. Since $M_1 \gg M_2$, the dynamics of inspiral is dependent on the mass ratio of the secondary to the mass of the DM particle \citep[e.g.,][chapter 8 and references therein]{Merritt2013degn.book.....M}. We define the mass-ratio of the secondary to the DM particle as $q_{\mathrm{DM}} \equiv \frac{m_{\mathrm{DM}}}{M_2}$. Three-body simulations show that using $q_{\mathrm{DM}} \leq 10^{-2}$, the results are convergent. The energy, angular momentum, and ejection time distributions are consistent between all of the models where $q_{\mathrm{DM}} \leq 10^{-2}$. Such a low mass ratio is desirable since it reduces any spurious three-body effects leading to sudden jumps in the evolution of the semi-major axis or eccentricity and ensures that the evolution is smooth. Accordingly, based on resolution and computational expenses, we choose $ m_{\mathrm{DM}} = 5\times10^{-5} M_{\odot}$ for the simulations using $\gamma_{\mathrm{sp}} = 7/3$ and $10^{-5} M_{\odot}$ when $\gamma_{\mathrm{sp}} = 9/4$. This results in $\approx 8\mathrm{k}-10\mathrm{k}$ particles in the simulation. Extra care is taken to ensure that the region surrounding the secondary have a sufficient number of particles to resolve the scattering process. Since the scattering process is quite stochastic, results from an individual $N$-body simulation are unreliable. Accordingly, five independent realizations of every model are generated and simulated.

For most of the simulations, only the $2.5PN$ term is added to the equation of motion of the binary. This is done in order to provide a direct comparison to previous studies such as \cite{Kavanagh2020PhRvD.102h3006K} and \citet{Becker2022PhRvD.105f3029B} where the effects of relativistic precession are neglected. In general, we  do not expect the dephasing to be affected by any precession effects. However, the net precession can itself be affected by the Newtonian precession of the binary in the potential of the spike in addition to the relativistic precession, which can also be a signature of the spike \citep[][]{Dai2022PhRvD.106f4003D}. It can, also, potentially affect the exchange of angular momentum between the spike to the IMRI. In order to understand if relativistic precession affects our results, we run a set of simulations with all PN terms enabled up to $2.5PN$ for $q=10^{-3}$ models. We leave a more systematic study of precession effects to future work. A summary of the initial conditions for the non-rotating models can be found in Table \ref{tab:summary_ic}.

We evolve the $q=10^{-2}, \mathrm{and} \; 10^{-3}$ models to merger. This is defined as the time when $a < r_{\mathrm{ISCO}}$. 
The simulations where $q=10^{-4}$ are extremely computationally expensive. We only achieve an evolution to 2.5 yr in 15-20 days of computing time. 
This results in about $2\times10^6$ orbits of the secondary around the primary. We calculate the dephasing cycles by extrapolating the rest of the evolution in vacuum to save on computational resources. To verify if the extrapolation produces valid results, we run a set of lower resolution simulations with 1024 DM particles to completion and compare our extrapolated results to the results from our lower resolution simulations. The extrapolated results can be seen as a lower limit on the number of dephasing cycles. We compare our results against the vacuum evolution computed using the \citet{Peter1964PhRv..136.1224P} analytic formula and also against those calculated using the Chandrasekhar DF assuming there is no backreaction on to the spike from the IMRI. This is done using the publicly available code {\tt\string IMRIPy} \citep{Becker2022PhRvD.105f3029B, Becker2023PhRvD.107h3003B, Imripy2023ascl.soft07018B}. The {\tt\string IMRIPy} results are denoted as ``Static DF" in the results section. Unfortunately, we could not use {\tt\string HaloFeedback} \citep{Kavanagh2020PhRvD.102h3006K} for our purposes since it can only simulate circular binaries.

\subsection{Rotating models}

As mentioned in the introduction, we aim to understand the effect of rotation in the DM spike and its effect on the IMRI. If the DM halo surrounding the IMBH is rotating, angular momentum conservation will dictate that DM spike should rotate as well. Additionally, spinning IMBHs will transfer angular momentum from the IMBH to the spike, torquing the spike up or down depending on the direction of the primitive rotation of the spike \citep{Ferrer2017PhRvD..96h3014F}. In a rotating spike, exchange of angular momentum between the IMRI and the spike will be enhanced compared to the non-rotating models and the eccentricity of the IMRI can be significantly affected. We aim to understand this effect in our study.
To the best of our knowledge, there is no literature available quantifying the level of rotation in the spike and its correlation to the galactic halo rotation or the spin of the central IMBH. Therefore, we caution the reader that our rotating models are somewhat ad hoc and therefore exploratory. 
However, they are still useful in understanding the qualitative effects a spike that is in a counter-rotating motion (retrograde motion) or co-rotating motion (prograde motion) with respect to the binary. 

To include rotation, we follow the Lynden-Bell trick \citep{LyndenBell1960MNRAS.120..204L}. This has been motivated by several studies which use the method to generate rotating models of galactic nuclei \citep[e.g.,][]{Boc2015ApJ...810..139H,Khan2020MNRAS.492..256K,Khan2021MNRAS.508.1174K,Var2021MNRAS.508.1533V} to examine the effects of rotation upon the dynamics of MBH binaries. In this method, we flip the $z$-component of the angular momentum of the DM particles, $L_{z,DM}$, to generate prograde or retrograde models. We note that the $z$-component of the angular momentum of the IMRI, $L_z$, is always positive. Accordingly, to generate co-rotating or prograde models, we reverse the direction of the $x$ and $y$ components of the velocity for all DM particles which satisfy $L_{z,DM} < 0$. To generate counter-rotating or retrograde models, we do the same, except with particles with $L_{z,DM} >0$. This leaves the density and velocity profiles of the spike unchanged.  In principle, this is not a fully self-consistent method of generating rotating models. Rotation is expected to flatten out the models to some extent so it is somewhat ad hoc to include rotation in spherical spikes. One can construct a distribution function $f(E,L_{z})$ as a function of the energy $E$ and $z$-component of the angular momentum $L_z$ as is done in \cite{Wang2014ApJ...780..164W} to generate self-consistent rotating models which are flattened. However, this is beyond the scope of this work. 
Irrespective of that, our models are able to qualitatively understand the effect of rotation on the IMRI. In future studies, we plan on investigating self-consistent models of rotating DM spikes to understand how the geometry of the spike affects the dynamics. 

We denote the ratio of retrograde particles to the total number of particles as $\mathcal{F}$. In our retrograde simulations, $\mathcal{F}=1$, indicating that \textit{all} particles are moving in a retrograde motion with respect to the binary, and in our prograde simulation, $\mathcal{F}=0$, which is opposite of the previous scenario. Realistic spikes are expected to have $0\leq \mathcal{F} \leq 1$. Since the simulations are quite expensive even with our fast $N$-body code, we restrict our study only to the $\mathcal{F}=0,1$ cases. This helps us put limits on the dephasing effect and compare the rotating models to fully isotropic non-rotating models. We note that the Lynden-Bell method is not the same as introducing rigid body rotation into the DM spike. The latter generates much stronger rotation whereas our method introduces a weaker level of rotation.

We generate prograde and retrograde rotating versions of all of the models presented in Table \ref{tab:summary_ic} and create five independent realizations for each configuration. All models were evolved to merger, except in the case where $M_1=10^4 M_{\odot}$ as explained in the section above.

\section{Results} \label{sec:results}
\begin{figure*} 
\centering
\begin{subfigure}[b]{1.0\textwidth}
\includegraphics[width=1.0\textwidth]{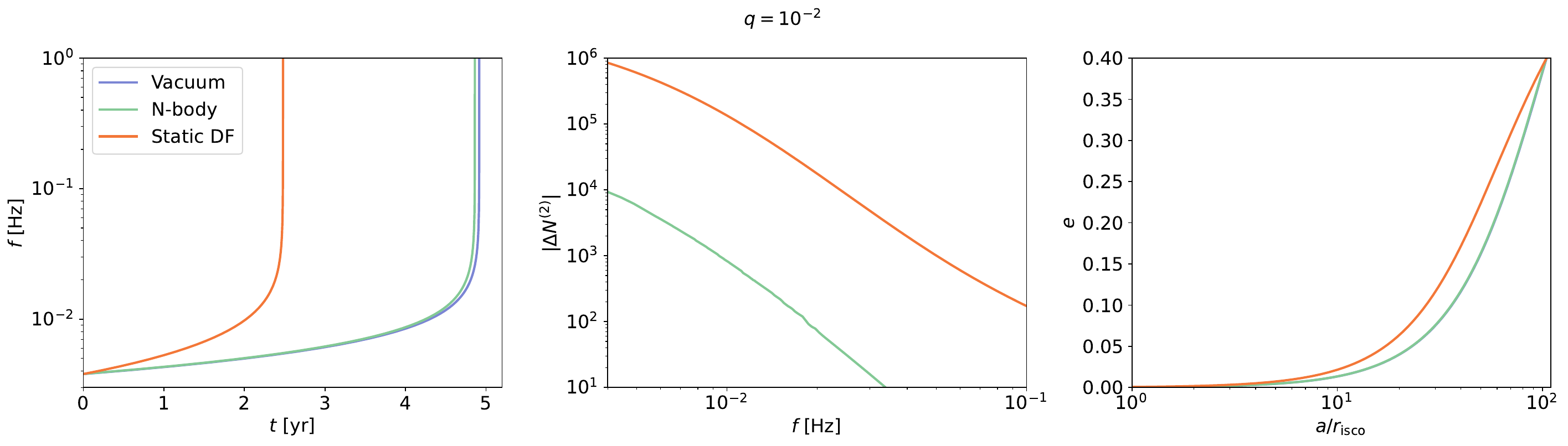}
\end{subfigure}
\begin{subfigure}[b]{1.0\textwidth}
\includegraphics[width=1.0\textwidth]{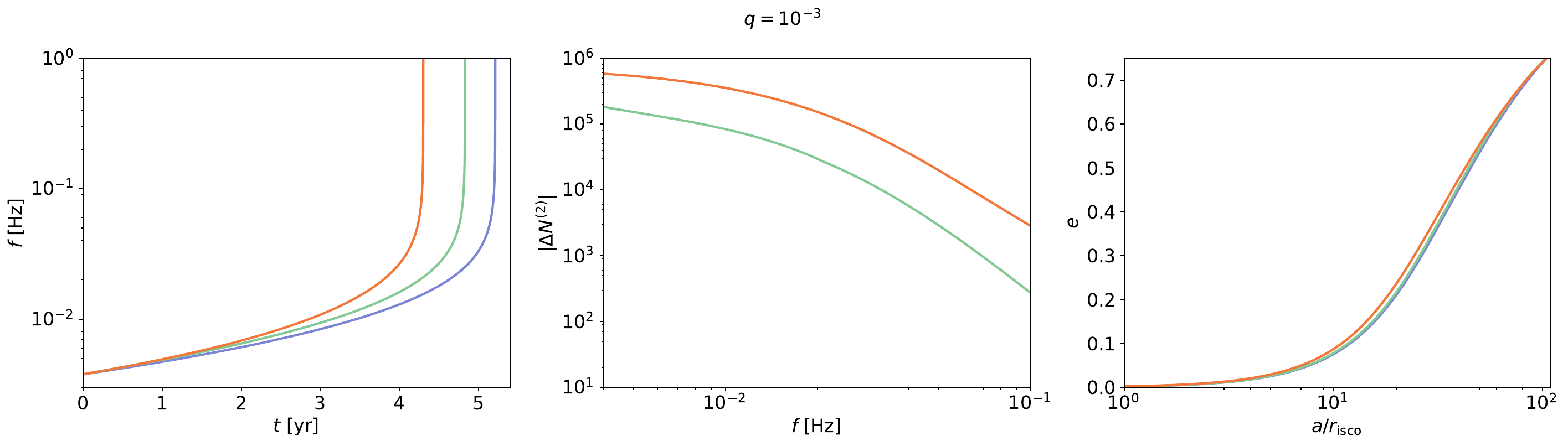}
\end{subfigure}
\begin{subfigure}[b]{1.0\textwidth}
\includegraphics[width=1.0\textwidth]{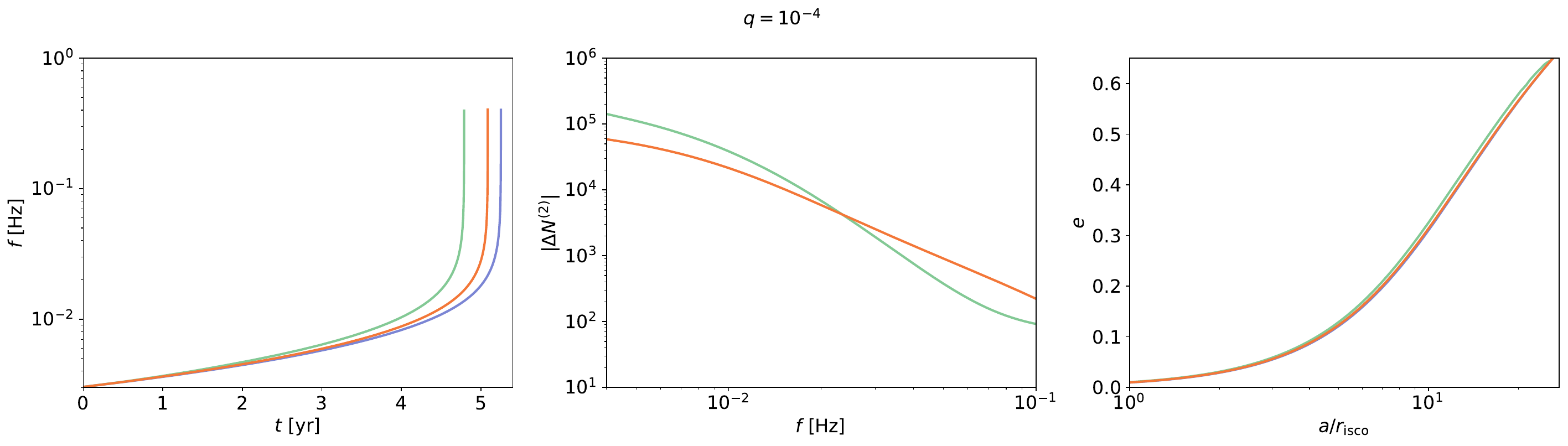}
\end{subfigure}
\caption{A comparison of the binary parameters for different mass ratio models evolving in a $\gamma_{\mathrm{sp}} = 7/3$ spike with $r_{\rm{sp}}$ calculated using equation \ref{eq:rsp}. The  evolution in vacuum (purple line) is calculated using \citet{Peter1964PhRv..136.1224P} analytic formula, while the evolution calculated using the Chandrasekhar DF formula assuming a static spike (orange line) is calculated using  {\tt\string IMRIPy}. They are compared to the evolution from our $N$-body simulations (green line). Left column: the mean orbital frequency $f$ as a function of time $t$ in years. Middle column: the estimated number of dephasing cycles of the second harmonic  $|\Delta N^{(2)}|$ as a function of the frequency $f$ in Hz. Right column: the eccentricity $e$ as a function of the semi-major axis $a$. We notice that in higher mass ratio models, the evolution of the binary is similar to that in vacuum. For $q=10^{-2}$ model, there is a $100\times$ reduction in the estimated number of dephasing cycles compared to the evolution calculated using Chandrasekhar DF formula as the spike has been disrupted in a very short time span. As we decrease the mass ratio, the disruption decreases. We notice that in case of the $q=10^{-3}$ model, the dephasing is only reduced by $3\times$ compared to the evolution calculated using {\tt\string IMRIPy}. For $q=10^{-4}$ model, we find that dephasing is a factor of 3 larger than what we obtain using the Chandrasekhar formula, with little to no disruption of the spike. This signals that the Chandrasekhar DF might be insufficient to explain the evolution of the binary in DM spikes. 
}
\label{fig:evolution_non_rot}
\end{figure*}

Owing to the inherent stochasticity present due to the discrete nature of $N$-body simulations, we note that the results presented here are calculated after taking the average of the quantities across all five independent realizations. In all of the calculated values, we noticed a maximum of 1 percent difference between results from individual simulations 
This does not affect our overall results and indicates the robustness of our simulations. We first present the results from the non-rotating models before moving on to the rotating models.

\subsection{Non-rotating models}

The mean orbital frequency of the binary $f$ can be written as 
\begin{gather} \label{eq:orbital_frequency}
    f = \frac{1}{2\pi} \sqrt{\frac{GM}{a^3}}
\end{gather}
where $M=M_1 + M_2$. 
We calculate the amount of dephasing by taking the difference between the number of GW cycles completed by the binary with and without the spike. 
Following \citet{Becker2022PhRvD.105f3029B}, we write the number of GW cycles $N^{(n)}(t_0,t_{\mathrm{final}})$ 
for the $n^{\mathrm{th}}$ harmonic between some initial time $t_0$ and final time $t_{\mathrm{final}}$ as follows:
\begin{gather}
    N^{(n)}(t_0,t_{\mathrm{final}}) = n \int_{t_0}^{t_{\mathrm{final}}} f (t) dt
\end{gather}
We can, then, calculate the difference in the number of GW cycles $\Delta N^{(n)} (t_0,t_{\mathrm{final}})$, or dephasing, as:
\begin{gather}
    \Delta N^{(n)} (t_0,t_{\mathrm{final}}) = N^{(n)}_{\mathrm{vacuum}} - N^{(n)}_{\mathrm{spike}}
\end{gather}
In our calculations, $t_{\mathrm{final}}$ is taken to be the time of merger 
For comparison amongst different models, we choose $n=2$, as has been the strategy in previous studies. Eccentric binaries radiate in multiple harmonics and the dephasing will be larger for larger $n$. However, these higher harmonics are harder to detect.

We present a comparison of the evolution of the mean orbital frequency of the binary $f$, the absolute value of number of dephasing cycles of the 2nd harmonic $|N^{(2)}|$, and the evolution of eccentricity of the binary $e$ as a function of its semi-major axis $a$ using $N$-body models and {\tt\string IMRIPy} for the non-rotating models in $\gamma_{\mathrm{sp}} = 7/3$ spike with $r_{\mathrm{sp}}$ calculated using equation \ref{eq:rsp} in Figure \ref{fig:evolution_non_rot}. Examining the $q=10^{-2}$ model, our $N$-body simulations predict that the binary merges 19 days earlier if it is embedded in a DM spike than in vacuum. It takes about $\approx 10^4$ fewer GW cycles 
than in vacuum to merge.  Comparing the $N$-body model to the DF model with a static spike, we find a $100\times$ reduction in the number of dephasing. The binary disrupts the spike almost completely within the first 0.1 years of inspiral leading to a drastic reduction in the amount of dephasing. This is unsurprising as the mass ratio of the binary is large which injects a substantial amount of energy ejecting DM particles around the secondary leading to the disruption of the spike. As such, the circularization rate of the binary is similar to that in vacuum as is evident from the $a-e$ plot.

\begin{figure}
    \begin{center}
    \includegraphics[width=0.45\textwidth]{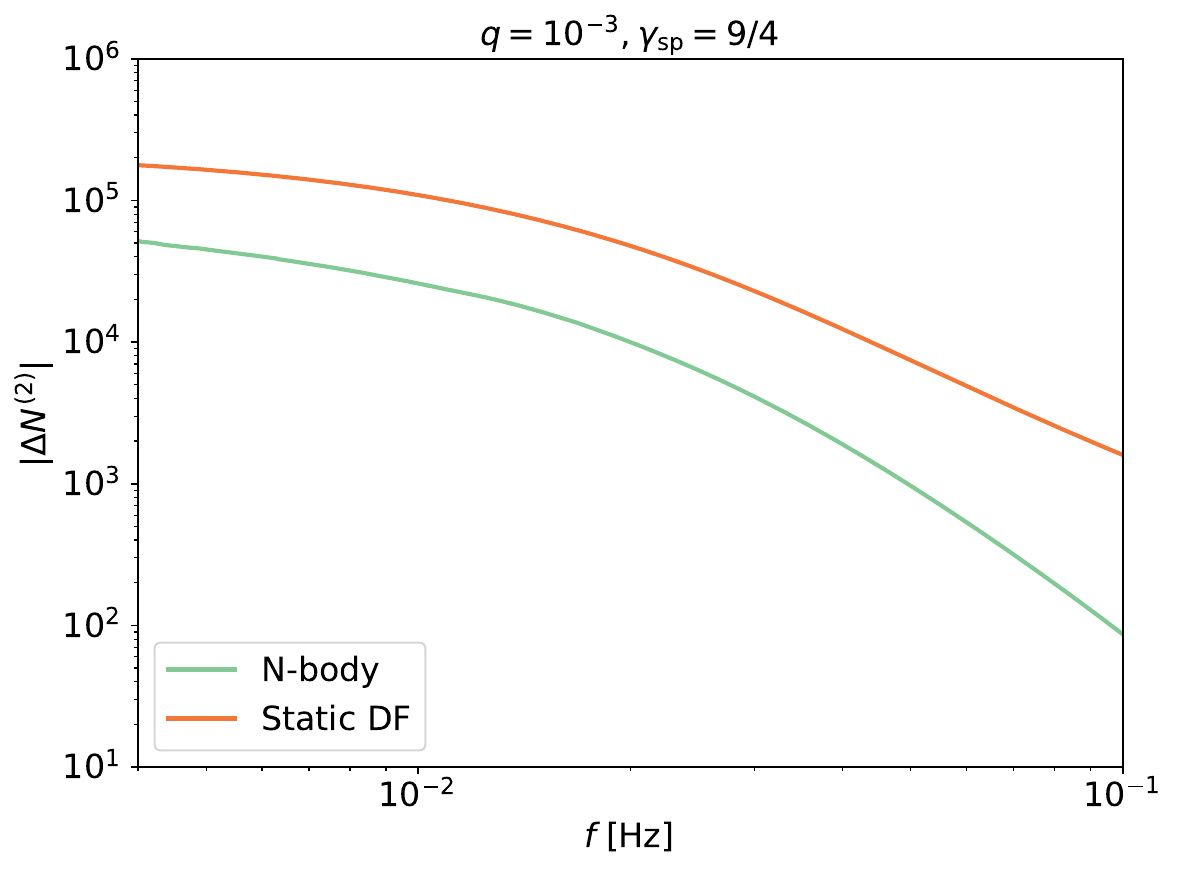}
    \caption{The dephasing of the second harmonic $|\Delta N^{(2)}|$ as a function of the binary frequency $f$ in Hz for a $q=10^{-3}$ binary embedded in a $\gamma_{\mathrm{sp}}=9/4$ spike. The color scheme is the same as that used in Figure \ref{fig:evolution_non_rot}. We notice that similar to the $\gamma_{\mathrm{sp}}=7/3$ case, the amount of dephasing in our $N$-body models is reduced by $\sim 3\times$ compared to the DF models.} 
    \label{fig:non_rot_2.25}
     \end{center}
\end{figure}

As we lower the mass ratio to $q=10^{-3}$, the binary merges earlier. It takes about 141 fewer days to merge in the $\gamma_{\mathrm{sp}}=7/3$ $N$-body model than its vacuum counterpart. This creates a larger amount of dephasing. In a $\gamma_{\mathrm{sp}}=7/3$ spike, the dephasing with respect to vacuum is $\approx 2\times 10^5$, only $3\times$ lower than what is predicted by the DF with static spike model. Although not directly comparable, we highlight the fact that this is much larger than the $100\times$ reduction for the same mass ratio found by \citet{Kavanagh2020PhRvD.102h3006K} in the semi-analytic {\tt\string HaloFeedback} models 
but with the secondary on circular orbit. Since the self gravity of the spike is minimal, we should expect a $\sim 3\times$ reduction in the amount of dephasing compared to the static DF models irrespective of the density profile. To verify that, we compare the dephasing cycles in the $\gamma=9/4$ model with $q=10^{-3}$ in Figure \ref{fig:non_rot_2.25}. We find that $|\Delta N^{(2)}| \approx 1.8 \times 10^5$ for the static DF models whereas  $|\Delta N^{(2)}| \approx 5.2 \times 10^4$, consistent with the $\sim 3\times$ reduction we expected. The rate of circularization is quite similar between the vacuum, static DF, and the $N$-body models. Nevertheless, we find that the eccentricity of the binary as a function of its semi-major axis for the $N$-body models lies approximately in between the static DF and the vacuum cases. 

\begin{figure}
    \begin{center}
    \includegraphics[width=0.45\textwidth]{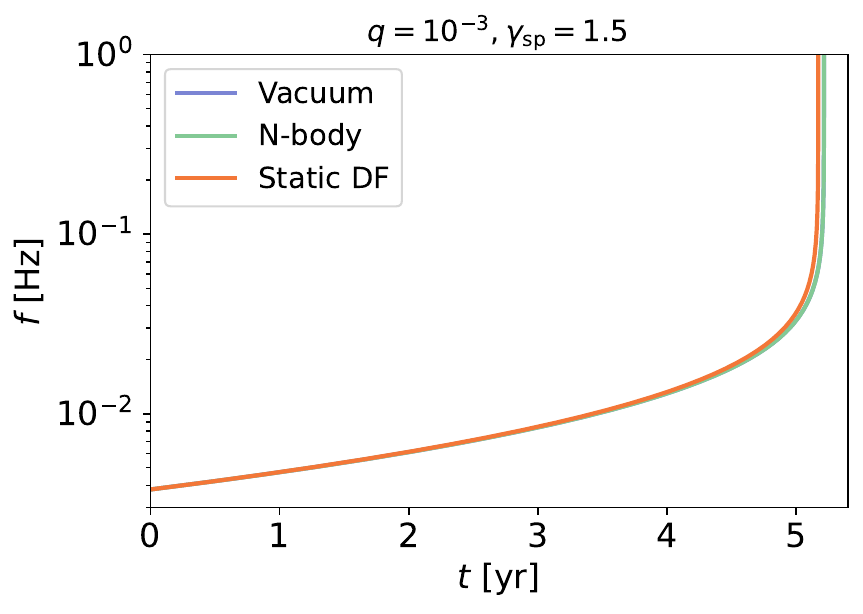}
    \caption{Binary frequency $f$ as a function of time $t$ for the $q=10^{-3}$, $\gamma_{\rm sp}=1.5$ model. We find that the low density of the spike results in no discernable differences between the inspiral in vacuum and that in the spike. This results in $\leq \mathcal{O}(10)$ cycles which might not be detectable. } 
    \label{fig:dm_crest_model}
     \end{center}
\end{figure}

For the $\gamma_{\rm sp}=1.5, q=10^{-3}$ case representing perturbed DM ``crests'', our findings are less optimistic. We note that such crests can also be formed due to adiabatic growth in an off-center IMBH \citep{ZhaoPhysRevLett.95.011301}. In Figure \ref{fig:dm_crest_model} we plot the evolution of the binary frequency $f$ as a function of time for this model. We find that the frequency evolution in the $N$-body models practically overlaps with that from the vacuum model. Unsurprisingly, the calculated dephasing is $\leq \mathcal{O}(10)$ which is larger than the limits of uncertainty from our $N$-body simulations. Such a small dephasing would also be much harder to detect than those from the denser spike models. Since the formation of the DM crests are more robust than that of DM spikes, our findings pose an important question of whether the spike models are overly optimistic. This is discussed further in the section \ref{subsec:realistic_spike}. 

\begin{figure}
    \begin{center}
    \includegraphics[width=0.45\textwidth]{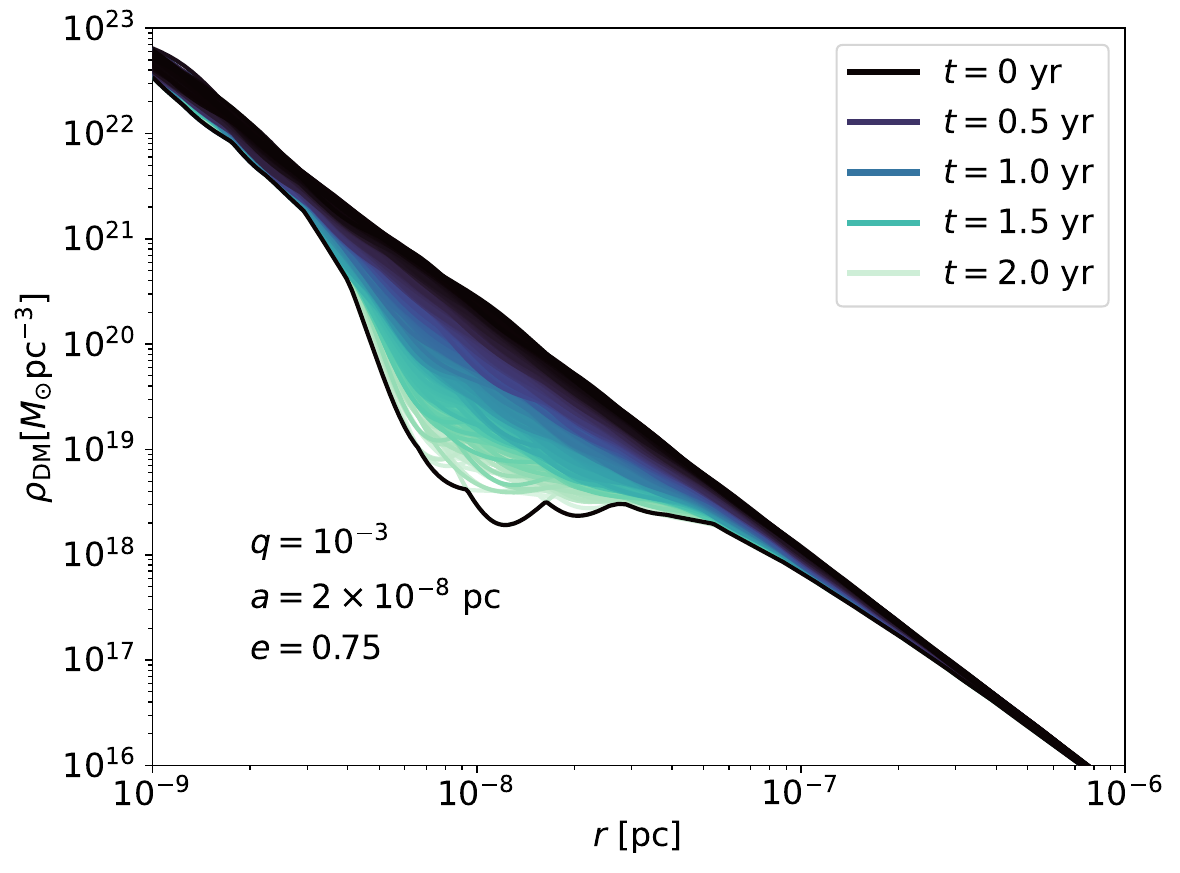}
    \caption{The density of the DM spike $\rho_{\mathrm{DM}}$ as a function of the distance $r$ from the primary in pc for the $q=10^{-3}$ model in a $\gamma_{\mathrm{sp}} = 7/3$ spike. The different colors represent the evolution of the density profile over time with darker colors representing earlier times and lighter colors representing later times. We notice that the inspiral of the IMRI leads to a significant disruption in the spike near the semi-major axis $a$ of the IMRI. The density of DM in that region is reduced by a factor of 100 compared to the original density after an inspiral time of 2 years. This is qualitatively consistent with the findings from \citet{Kavanagh2020PhRvD.102h3006K}. However, we find that the disruption to the spike due to the IMRI in our $N$-body is much slower than what is found by \citet{Kavanagh2020PhRvD.102h3006K} who find that the spike is significantly disrupted within 0.1 years. } 
    \label{fig:density_evol}
     \end{center}
\end{figure}

We examine the evolution of the density profile in the $q=10^{-3}$, $\gamma_{\mathrm{sp}}=7/3$ model over time due to the effect of the binary in Figure \ref{fig:density_evol}. The binary injects energy into the spike by scattering the DM particles, thereby reducing the density of the spike.
It preferentially interacts and scatters particles near itself, so we expect a drop in the density profile near the binary over time. The local efficiency of scattering will depend on the orbital parameters of the binary and density profile near the secondary.  During a 2 year time span, we notice that the binary has drastic effects on the density of the spike. There is up to a factor of 100 reduction in the density of DM particles near the initial semi-major axis of $a=2\times10^{-8}$. The density of DM particles reduces from $\sim 10^{20} M_{\odot} \mathrm{pc}^{-3}$ to $\sim 10^{18} M_{\odot} \mathrm{pc}^{-3}$. The binary effectively carves out a flat core near its semi-major axis. Since the secondary spends a larger time near its apoapsis than its periapsis, we expect the scattering to be most effective near the apoapsis carving out a core. The apoapsis of the secondary is $3.75 \times 10^{-8}$ pc. We notice that the flat core extends from $\sim 10^{-8}$ pc to $4\times10^{-8}$ pc consistent with the hypothesis. The impact further away, and near the periapsis is smaller. This has important implications for the survival and detectability of DM spikes and is discussed at length in section \ref{subsec:realistic_spike}.  Similar findings were noted in \citet{Kavanagh2020PhRvD.102h3006K} in case of circular binaries. However, \citet{Kavanagh2020PhRvD.102h3006K} find that the disruption happens quite quickly, effectively within 0.1 yr from the beginning of the simulation for $q=10^{-3}$, whereas in our case the disruption is much slower, by almost $20\times$. This leads to a much larger dephasing effect compared to the {\tt\string HaloFeedback} models. Notably, the evolution of the density profile does not show the presence of a wake or a density ``bump" behind the secondary, which is present in the self-consistent {\tt\string HaloFeedback} models. According to the authors, the ``bump" is caused due to scattering of DM particles near the secondary which give rise to the DF effect. The absence of the ``bump" and the longer spike disruption time suggests that DF theory is inconsistent with the results from our $N$-body simulations. Additionally, we find that the effect of softening is minimal on the feedback time and effects on the density profile. When a softening of $\approx 10^{-10}$ pc is used between the secondary and DM particles, we find that the feedback at $r < 10^{-8}$ pc is decreased but the differences between the density profiles in that region between non-softened and the softened sims never vary substantially.

\begin{figure}
    \begin{center}
    \includegraphics[width=0.45\textwidth]{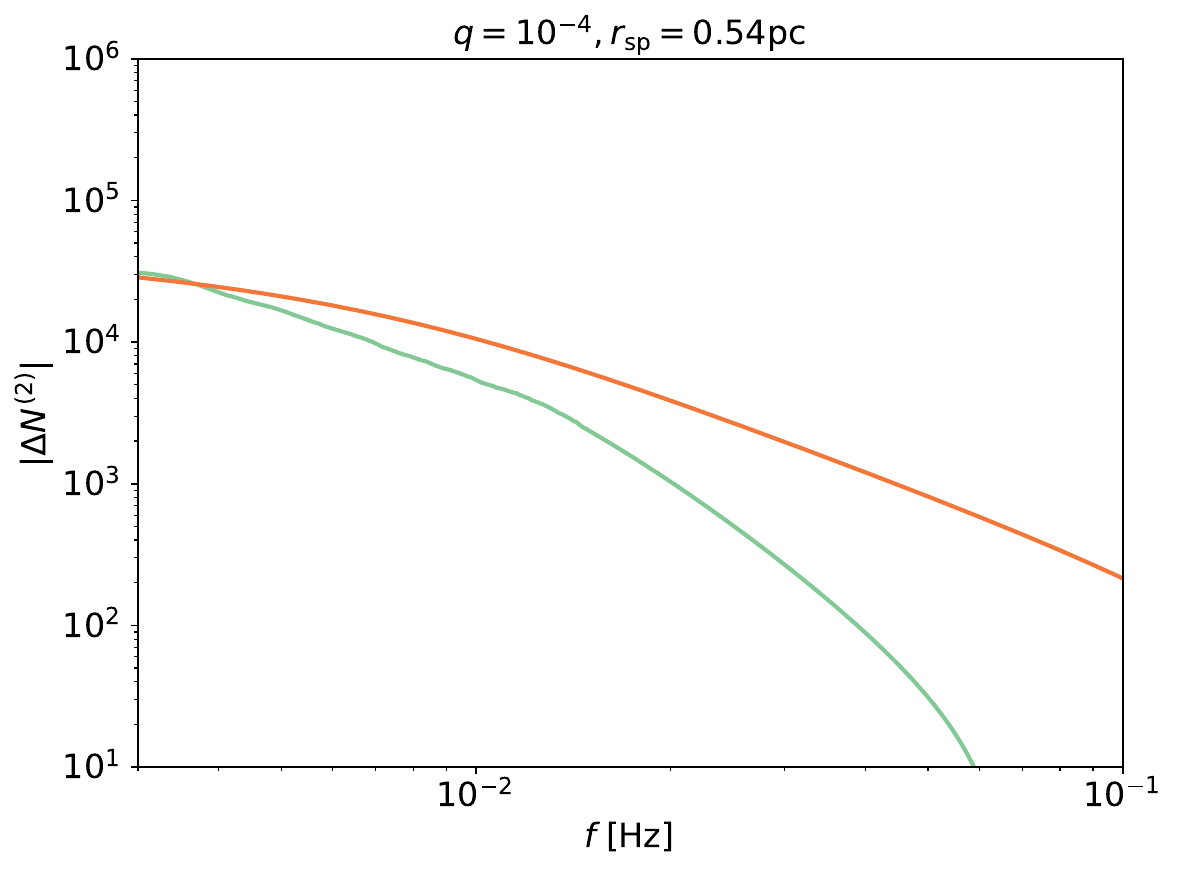}
    \caption{Similar to  Figure \ref{fig:non_rot_2.25} but for $q=10^{-4}$ binary embedded in a $\gamma_{\rm{sp}}=7/3$, $r_{\rm{sp}}=0.54 \rm{pc}$ spike. We notice that the net dephasing predicted by the $N$-body simulation is comparable to that predicted by the {\tt\string IMRIPy simulation}, although at higher frequencies the dephasing falls off faster in the $N$-body models. The density in the region of interest for this particular model is about 6 times lower than that for the $q=10^{-4}$, $r_{\rm{sp}}=1.17 \rm{pc}$ model and we notice a similar decrease in the amount of dephasing in this scenario compared to that model. } 
    \label{fig:non_rot_q2_lower}
     \end{center}
\end{figure}

For our $q=10^{-4}$ non-rotating model with $r_{\rm sp}=1.17$ pc, examining Figure \ref{fig:evolution_non_rot}, we notice a very surprising result. While the binary in the static DF model merges 60.3 days faster then the vacuum case, the binary in the $N$-body model merges 169 days faster. This leads to dephasing effects that are almost $3\times$ larger than the DF model. Whereas in the static DF case we find that $|\Delta N^{(2)}| \approx 6\times10^4$, our $N$-body simulation suggests that $|\Delta N^{(2)}| \approx 1.5\times10^5$. Since the results are extrapolated beyond 2.5 years using the \citet{Peter1964PhRv..136.1224P} equation, we compared them to those obtained from set of lower resolution simulations which are run to completion. We find that the extrapolated results are consistent with those from the lower resolution simulations. The extrapolated results fall within a standard deviation of the results from the lower resolution simulations.  For the models where we fix $r_{\rm{sp}}=0.54$ pc, similar to the $q=10^{-2}$ and $10^{-3}$ models, we notice a decrease in the amount of dephasing compared to the $r_{\rm{sp}}=1.17$ pc model. We present the dephasing in $q=10^{-4}$, $r_{\rm{sp}}=0.54$ pc model in Figure \ref{fig:non_rot_q2_lower} and find that the $N$-body and the {\tt\string IMRIPy} simulations predict a similar amount of dephasing, with $|\Delta N^{(2)}|=3\times10^4$. This is $\approx 6\times$ lower than that obtained in the $r_{\rm{sp}}=1.17$ pc models which is unsurprising given the lower density in the  $r_{\rm{sp}}=0.54$ pc models. The density in the region of interest is $6.1 \times$ lower in the model with $r_{\rm{sp}}=0.54$ pc compared to the model with $r_{\rm{sp}}=1.17$ pc, which explains the factor of 6 reduction in the dephasing. This, along with the results from the $q=10^{-3}$ models, strongly suggests that the dephasing is proportional to the local density of the profile at the position of the secondary. Similar to our findings from the  $q=10^{-3}$ model, we find the spike plays a very little role in dephasing if $\gamma_{\rm sp}=1.5$ even when $M_1=10^4 M_{\odot}$. The dephasing is again $\leq \mathcal{O}(10)$, and below our uncertainty thresholds.

The results obtained for the $r_{\rm{sp}}=1.17$ pc case for the $q=10^{-4}$ binary is a factor of 15 greater than what was found by \citet{Kavanagh2020PhRvD.102h3006K} in their {\tt\string HaloFeedback} simulations for similar models on circular orbits. We also compare the spike density profiles at the beginning and end of our simulation and find that there is no degradation in the density profile of the spike, while the reduction of the DM density is observed in the {\tt\string HaloFeedback} models even with such low mass ratios. This, along with our previous results, suggests that previous analytic/semi-analytic calculations using DF underestimated the total number of dephasing cycles by a factor of one to ten (or even larger) for lower-mass ratio binaries. This is especially evident while comparing the $N$-body simulations to the {\tt\string HaloFeedback} simulations, where we find a difference of ten to hundred times. We use three-body simulations in section \ref{subsec:three_body_sims} to show that three-body scattering provides a better description of the dynamics of the binary in the spike and could explain the inconsistency observed in this section, while showing that the impact of DF is subdominant.

\subsection{Rotating models}

We now move on to the analysis of the binary inspiral in the rotating models. From DF theory, we would expect the inspiral of the binary in counter-rotating or retrograde models to be slower than that in co-rotating or prograde models. This is caused due to the fact that in the prograde models, the relative velocity of the DM particles is smaller than that in the retrograde models. Since the force of DF is inversely proportional to the relative velocity of the interacting particles, we would expect the DF force to be larger in the prograde scenario, resulting in a faster inspiral. 

\begin{figure*} 
\centering
\begin{subfigure}[b]{1.0\textwidth}
\includegraphics[width=1.0\textwidth]{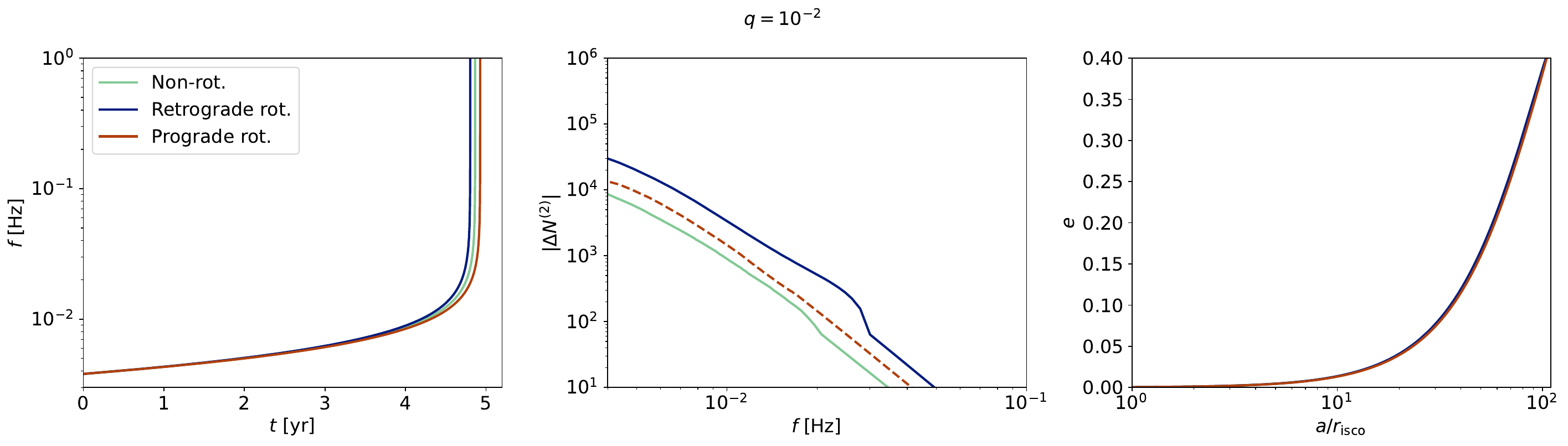}
\end{subfigure}
\begin{subfigure}[b]{1.0\textwidth}
\includegraphics[width=1.0\textwidth]{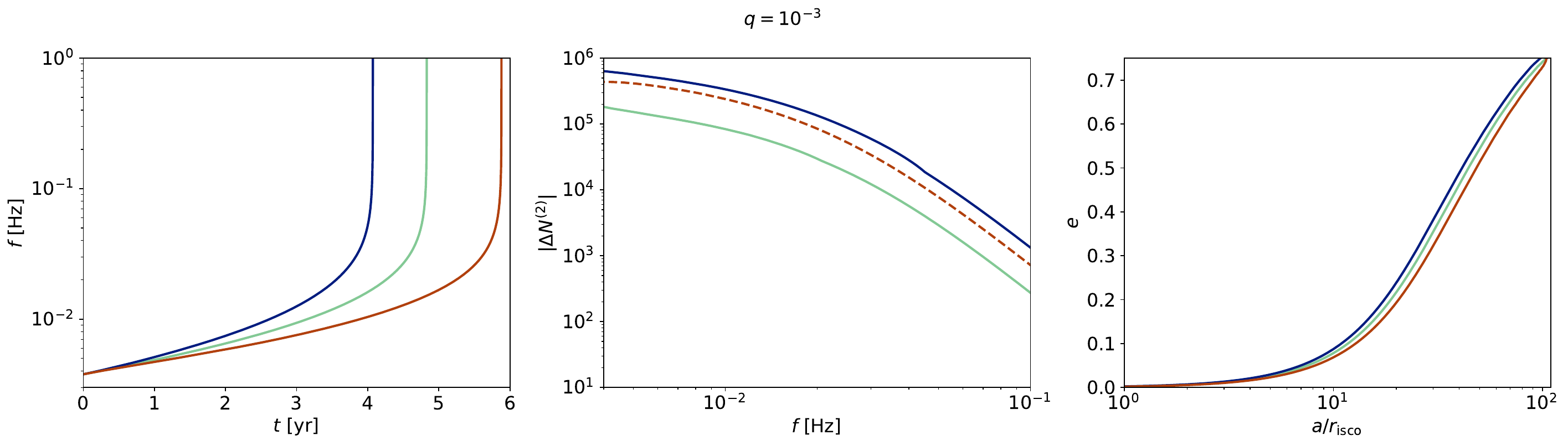}
\end{subfigure}
\begin{subfigure}[b]{1.0\textwidth}
\includegraphics[width=1.0\textwidth]{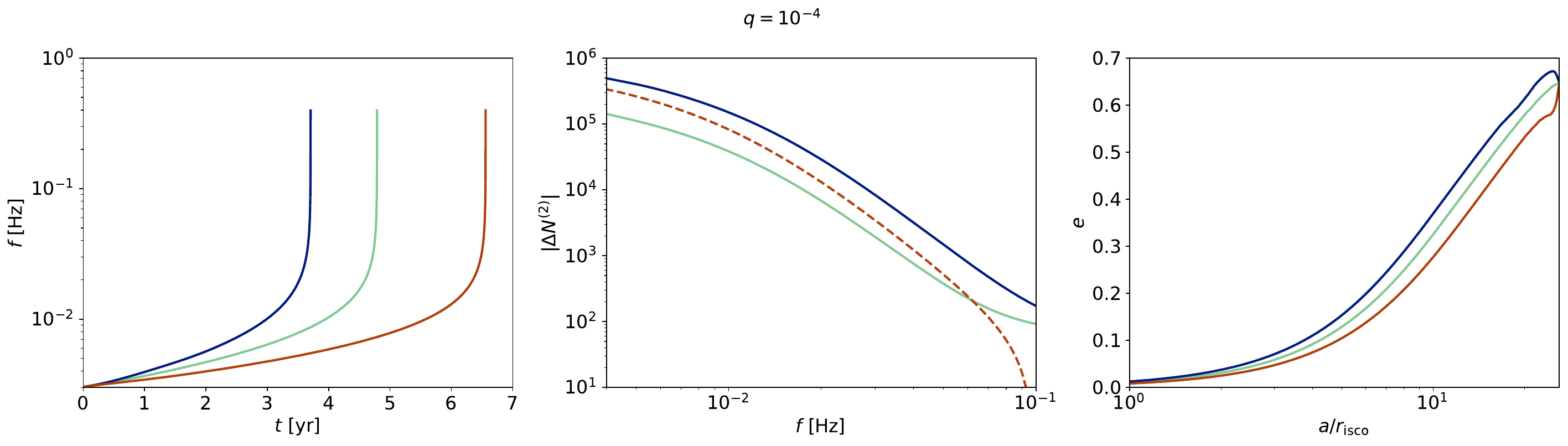}
\end{subfigure}
\caption{A comparison of the binary evolution parameters for different mass ratio models inspiraling in a $\gamma_{\mathrm{sp}} = 7/3$ spike similar to Figure \ref{fig:evolution_non_rot} but also including the evolution from rotating models. We notice that the in all of the retrograde models, the binary merges faster than in the prograde and even the non-rotating models. This leads to a major enhancement in the number of dephasing cycles, by as much as $3\times$ that of the non-rotating model in the case of $q=10^{-4}$. Since we are plotting the absolute value of the number of dephasing cycles, the dephasing in case of the prograde models appears to be positive. However, the prograde models actually merge slower than even the vacuum models and $\Delta N^{(2)}$ is actually \textit{negative} (indicated by the dashed lines) . Thus the number of GW cycles in prograde models is larger than that in vacuum.  We also find that in the retrograde models, the binary circularization rate is slower than that in the prograde models. In the latter scenario, the circularization is enhanced leading to a slower inspiral. As the mass ratio decreases, the effect of rotation becomes more prominent suggesting that transfer of angular momentum between the spike and the IMRI contributes majorly and cannot be ignored. }
\label{fig:evolution_rot}

\end{figure*}

Examining the properties of the binary over time in both the prograde models and retrograde models in Figure \ref{fig:evolution_rot} we notice a very surprising result. We find that in all of the cases, the retrograde models merge faster than the prograde and even the non-rotating models. Even more astonishing is the fact that the prograde models are slower to inspiral than the vacuum models. This is the opposite of what we would have expected from DF.
Examining the $q=10^{-2}$ models, we find that the expected amount of dephasing increases by almost $4\times$ compared to its non-rotating counterpart. Since we are plotting the abolute value of the number of dephasing cycles with respect to vacuum, we find that the prograde model has a larger amount of dephasing, almost $1.5\times$ that of the non-rotating model. However, we want to stress that the prograde model actually take longer to inspiral than the vacuum model. So it actually takes $1.5\times10^4$ more GW cycles than the vaccum model to merge. 

As we lower the mass ratio, we notice the same trend. In the $q=10^{-3}$ models, we find that $|\Delta N^{(2)}|\approx 7\times 10^5$ in the retrograde model, whereas $|\Delta N^{(2)}|\approx 2\times 10^5$ in the non-rotating model, almost $3.5\times$ lower. Similar to the $10^{-2}$ model, we find that the prograde model takes $4.5\times10^5$ more GW cycles to merge than the vacuum model. Interestingly, we also notice that the retrograde model eccentrifies quickly in the beginning, before circularizing later due to GW effects. The opposite is observed in prograde models. There is an accelerated circularization in the beginning, followed by a period of circularization led by the emission of GW. 

As we lower the mass-ratio to $q=10^{-4}$, we notice that the eccentrification and circularization effects increase. Examining the eccentricity as a function of the semi-major axis, we find that the evolution in the retrograde and prograde models diverge quickly. The binary eccentrifies quickly, reaching a maximum eccentricity of 0.67 before circularizing. The opposite happens in the prograde models where the binary circularizes quickly, within the first one year time span, reaching an eccentricity of 0.61. This eccentrification and circularization lead to accelerated or deccelerated inspiral due to GW emission since the GW emission is very sensitive to the binary eccentricity. This leads to the larger dephasing we see in retrograde models. We find that the retrograde model merges about 550 days earlier compared to the vacuum model, whereas the prograde model takes 484 days longer. In the retrograde models, this produces a dephasing effect that is 2.5 times as large as the non-rotating models, with the binary taking $5\times10^5$ fewer cycles to merge compared to the vacuum models. The opposite effect is noted in the prograde models with the binary taking $3.5\times10^5$ more cycles to merge. Interestingly, in all of our simulations we find that the ratio of dephasing cycles in retrograde rotating to non-rotating models is between $2.5-3$. Similarly for prograde models, we find that the ratio of dephasing cycles in rotating to non-rotating models is about $1.5-2$, independent of the density profile and mass ratio. This is likely to be an artifact of our chosen initial conditions rather than a fundamental property of rotation itself. 

We also compare the density profiles of the non-rotating to the rotating models over time to understand if the feedback from the binary to the spike changes upon the inclusion of rotation. Surprisingly, we find that the density profiles are consistent among the rotating and non-rotating models. Although rotation changes the dynamics of the binary significantly, the effect of binary on the spike is consistent among non-rotating and rotating models.

\subsection{Three-body simulations} \label{subsec:three_body_sims}

The results obtained in the previous section suggest that DF theory is insufficient at explaining the long term dynamics of the binary. In fact, this is not a very surprising result as it has been known that upon the formation of a hard binary, three- body scattering becomes more effective than DF at dissipating energy, similar to SMBH binaries \citep[e.g.,][]{Begelman1980Natur.287..307B,Merritt2013degn.book.....M}.  
A binary is said to become a hard binary when the separation between the primary and secondary falls below the hard binary radius $a_{\rm h}$ which is defined as
\begin{gather}
      a_{\rm h} = \frac{q}{(1+q)^2} \frac{d_{\rm infl}}{4},
\end{gather}
where $d_{\rm infl}$ is the influence radius of the primary and is defined as the radius that encloses twice the mass of the primary \cite[e.g.,][]{Mukherjee2023MNRAS.518.4801M}. For our models, we find that $a_{\rm h}$ ranges from $6.75\times10^{-3} \mathrm{pc}$ in the $q=10^{-2}$ model to $1.46\times10^{-4} \mathrm{pc}$ in the $q=10^{-4}$ model. Therefore, all of our binaries are in the hard binary limit at the beginning of our simulations. 

In the three-body scattering 
 scenario, the binary undergoes a complicated three-body interaction in which the incoming particle interacts with the binary multiple times. This results in the ejection of the particle eventually, leading to shrinkage of the binary's orbit. This process has been called the \textit{slingshot mechanism} and is fundamentally different from Chandrasekhar DF which is a  cumulative effect of two-body encounters. 

 To understand which of the two above-mentioned processes dominate, we can analyze the $N$-body simulations and compare the efficiency of three-body scattering to that of dynamical friction. The DM particles are orbiting the primary and are bound to it initially. This allows us to obtain the semi-major axis $a_{\mathrm{DM}}$ and eccentricity $e_{\mathrm{DM}}$ of each particle. From our $q=10^{-3}$ model with $\gamma_{\rm{sp}}=7/3$, we select a subset of DM particles which satisfy $0.5 a_0 \leq a_{\mathrm{DM}} \leq 1.5 a_0$, where $a_0$ is the initial semi-major axis of the binary. In principle, all particles interact with the binary, but particles with semi-major axis close to the binary's interact strongly. We, then, calculate the initial and final energies of the selected particles at the beginning and end of our simulations to calculate the change in energy of the DM particle ($\Delta E_{\rm DM}$). We find that all of the selected particles are ejected from the spike by the end of the simulation.
 
 The selected particles are evolved from their initial positions along with the binary individually using  {\tt\string IAS15} for the same duration as the full $N$-body simulations. The energy of each particle is recorded at the beginning and end of the simulation. The difference in energies provides an estimate of the change in energy due to three-body effects.
 On the other hand, calculating the energy dissipation due to dynamical friction from the $N$-body simulations is somewhat non-trivial. According to \cite{Chandra1943ApJ....97..255C}, as $M_2$ moves through the medium of DM particles, it experiences a number of two-body encounters that change the velocity of the secondary in a direction parallel and opposite to the initial velocity of the secondary. Each particle contributes a net change in the velocity of the secondary giving rise to dynamical friction force over time. We can estimate the dynamical friction force due to each individual particle using the method described in \citet{Ma2023MNRAS.519.5543M}.  The energy change due to dynamical friction attributed to the $i^{\rm th}$ DM particle $\dot{E}_{\mathrm{i,df}}$ can be written as  
\begin{gather}
    \dot{E}_{\mathrm{i,df}} = M_2 \mathbf{a}_{\mathrm{i,df}} \cdot \mathbf{v}_2
\end{gather}
where $\mathbf{a}_{\mathrm{i,df}}$ can be calculated using equation 9 from \citet{Ma2023MNRAS.519.5543M} and $\mathbf{v}_2$ is the velocity of the secondary. We calculate $\dot{E}_{\mathrm{i,df}}$ using the saved snapshots and integrate over time to find the net energy change due to dynamical friction. The results from our simulations are stored at a fine enough time resolution that this method is able to provide a good approximation of the dynamical friction energy loss.

\begin{figure}
    \begin{center}
    \includegraphics[width=0.45\textwidth]{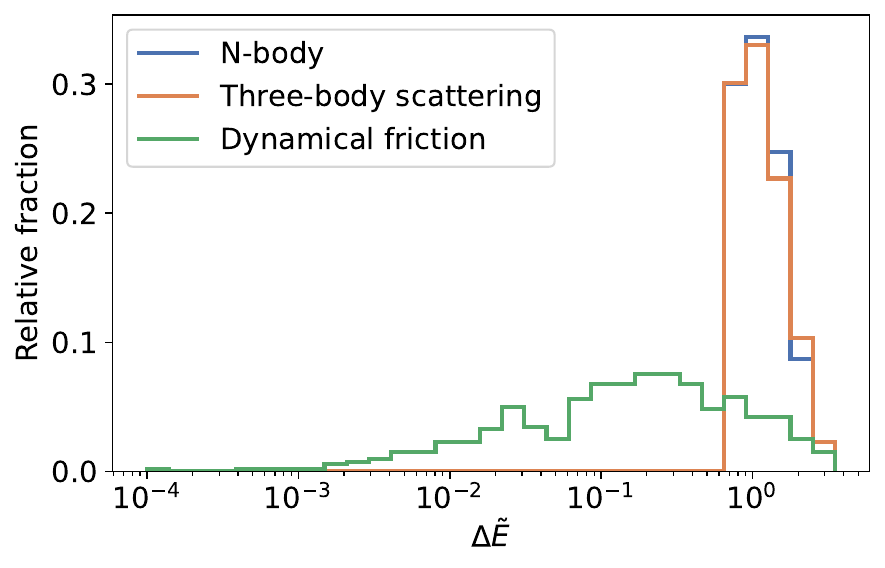}
    \end{center}
    \caption{The distribution of the normalized energy change $\Delta \tilde{E}$  of strongly interacting DM particles from our $N$-body simulations compared to that from three-body scattering simulations and  dynamical friction approximation. We find that the distribution of energy change of strongly interacting particles is consistent between the $N$-body simulations and the three-body simulations but not with the calculated values using the dynamical friction approximation. This confirms our hypothesis that three-body scattering, and not dynamical friction is responsible for dissipating energy from the binary.}
    \label{fig:threebody_versus_df}
     
\end{figure}

We present the relative distribution of the energy change of the selected particles from the $N$-body simulation, and compare it to the energy change due to three-body effects and dynamical friction in Figure \ref{fig:threebody_versus_df}. For clarity, we present the normalized change in energy $\Delta \tilde{E}$ where we normalize the change in energy with respect to the binding energy of the DM particle with $a_{\mathrm{DM}}=a_0$. We find that the distribution of the change in energy from the $N$-body simulations matches that from the three-body simulations but is inconsistent with the energy change due to dynamical friction.
The $N$-body and three-body simulations show that each selected particle experiences, on average, a normalized energy change of 1.0-1.5. In the dynamical friction scenario, the average energy loss per particle is about $10\times$ lower. Additionally, we find that, akin to the $N$-body simulations, all particles are ejected from the system in the three-body simulations. From the estimates of the dynamical friction force, we find that only 10-20 per cent of the particles would be ejected from the system entirely, consistent with the findings from \citet{Kavanagh2020PhRvD.102h3006K}, but is in tension with the findings from the $N$-body simulation.
This confirms our hypothesis that three-body scattering, and not dynamical friction is the predominant method of energy dissipation in our simulations. Although not presented here, we find a similar story across all of the models used in this study. Interestingly, as the mass ratio is decreased, the fraction of energy loss due to the three-body scattering increases. This is consistent with \citet{Merritt2013degn.book.....M} where the three-body scattering efficiency is proportional to $q^{-1}$.  
We find a similar distribution in the change in energy of the particles from both the three-body and full $N$-body simulations in our $q={10}^{-4}$ models as well. Our results also highlight that our $N$-body simulations are robust, accurate, and consistent with results from the extremely accurate {\tt\string IAS15} integrator where the net energy is conserved to machine precision.

What causes the counter intuitive results from our rotating simulations? \citet{Merritt2009ApJ...693L..35M}, \citet{Iwasawa2011ApJ...731L...9I}, and \cite{Sesana2011MNRAS.415L..35S} hold clues that are able to shed some light on this mystery. During a three-body encounter, the binary exchanges energy and angular momentum with the particle in a complicated fashion. \citet{Iwasawa2011ApJ...731L...9I}, in mergers of MBH binaries in galactic nuclei, noted that counter-rotating stars are much more effective in extracting angular momentum from the binary during the three-body scattering phase. As explained by \citet{Merritt2009ApJ...693L..35M} and \citet{Sesana2011MNRAS.415L..35S}, this is caused due to the torquing mechanism of the binary's potential on to the particle which leads to a secular evolution of the particle's eccentricity and inclination. This mechanism converts the counter-rotating particles to co-rotating which are then preferentially ejected by the binary \citep{Iwasawa2011ApJ...731L...9I}. This results in a larger change in angular momentum of the particle and as such, counter-rotating particles are able to extract angular momentum from the binary more efficiently compared to co-rotating particles. This process becomes more efficient for more eccentric binaries.

To understand if the mechanism mentioned above is able to explain the counter-intuitive results from the previous sections and to provide a better description of the dynamics of the binary, we perform three-body simulations. This is done to understand the transfer of energy and angular momentum during the scattering of a DM particle by the binary. We follow the similar steps as in \citet{Sesana2011MNRAS.415L..35S} to set up our simulation with some differences. 

\begin{figure*}
    \begin{center}
    \includegraphics[width=1.0\textwidth]{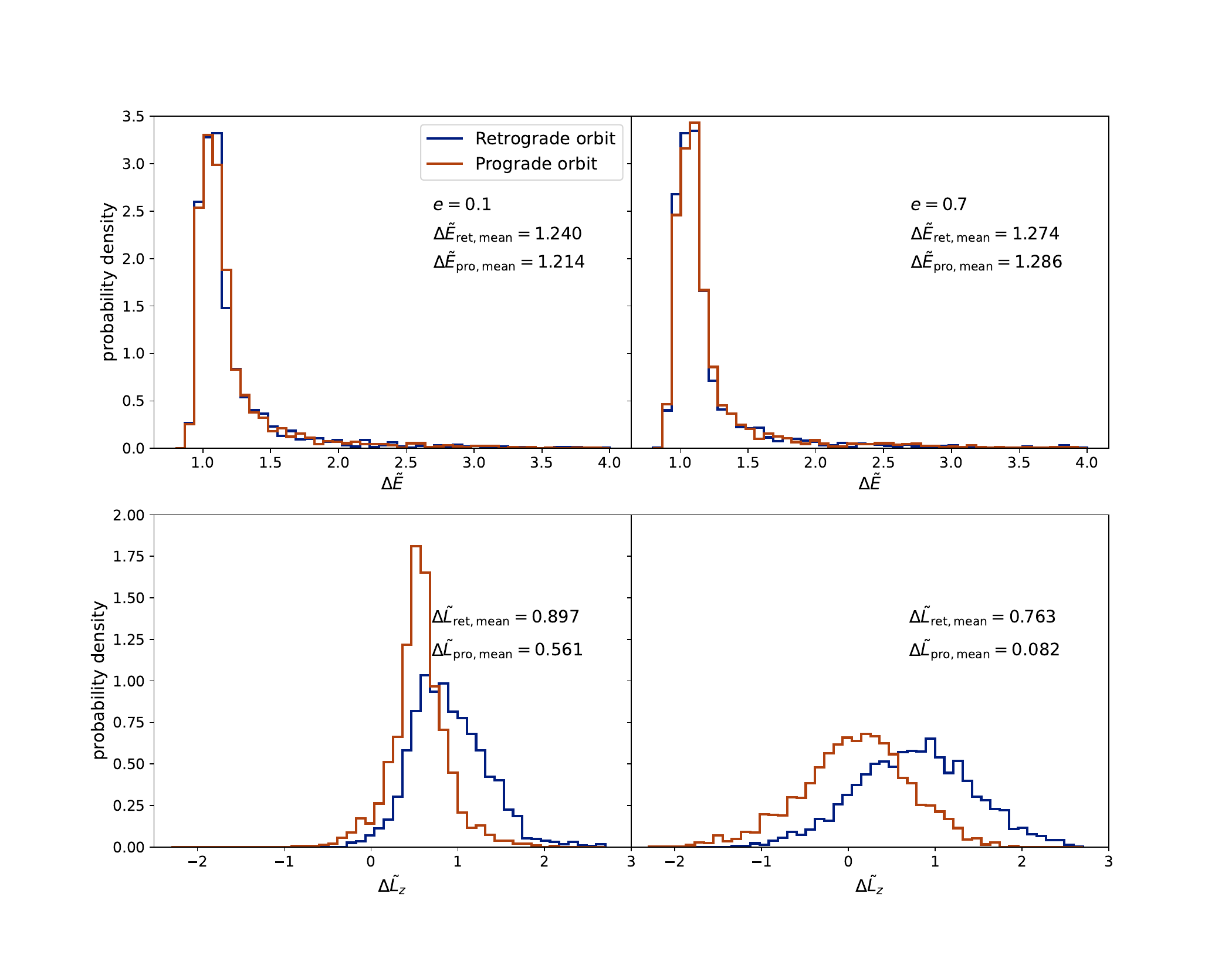}
    \caption{Top: probability distribution of the normalized energy change $\Delta \tilde{E}$ of $2500$ DM particles from the three-body simulations performed until the particle is ejected by interactions with the binary for both prograde and retrograde rotation models with two different binary eccentricities $e$. We notice that in both prograde and retrograde models, the distribution of the normalized energy is similar and does not change with the binary eccentricity. Bottom: similar to top but the probability distribution of the normalized angular momentum $\Delta \tilde{L_z}$ of $2500$ DM particles after ejection. We notice that there are major differences in the distribution of $\Delta \tilde{L_z}$  between prograde and retrograde models. In particular, in the retrograde models, the angular momentum of the particle is larger than that in the prograde models, especially for more eccentric binaries. This suggests that retrograde particles are more efficient at ``stealing'' angular momentum from the binary, especially at larger $e$.    } 
    \label{fig:three_energy_mom_hist}
     \end{center}
\end{figure*}

We first generate a $N=1M$ particle realization using {\tt\string Agama} containg a primary with mass $M_1 = 10^{3} M_{\odot}$ and $\gamma_{\mathrm{sp}}=7/3$ and then use the Lynden-Bell trick to generate prograde and retrograde models. We then place the secondary with mass $M_2 = 10 M_{\odot}$ at $a_0=2\times10^{-8}$ pc with varying eccentricity. We ensure that the binary lies in the $x-y$ plane. From the generated initial model, we only select DM particles which have $a_{\mathrm{DM}} \approx a_0$. This results in $\sim 2500$ DM particles being selected. We, then, evolve each particle with the binary individually, running 2500 three-body simulations using {\tt\string IAS15}, until the DM particle is completely ejected from the system. We record the initial and final values of the energy and angular momentum of the particle and the binary. We note that the simulations are purely Newtonian.

The eccentricity of the binary can be written as follows:
\begin{gather} \label{eq:sesna01}
    e = \sqrt{1-\frac{2EL^2}{G M^2 \mu^3}} = \sqrt{1-\frac{2EL_{z}^2}{G M^2 \mu^3}}
\end{gather}
where $E$ is the binding energy of the binary, $L$ is the angular momentum of the binary, $M=M_1+M_2$, and $\mu = \frac{M_1 M_2}{M_1 + M_2}$. The last equality follows from the fact that our binary is in the $x-y$ plane. As such, $L = L_z$. 
Differentiating equation \ref{eq:sesna01} as is done in \citet{Sesana2011MNRAS.415L..35S} provides us with the change in eccentricity $\Delta e$, which we find to be:
\begin{gather} \label{eq:sesana02}
    \Delta e = -\frac{1-e^2}{2e} \left ( \frac{\Delta E}{E} + \frac{2 \Delta L_z}{L_z} \right) = \frac{1-e^2}{2e} \chi 
\end{gather}
where $\Delta E$ is the change in $E$, $\Delta L_z$ is the change in $L_z$ and  $\chi$, called the eccentrification parameter, has been defined using the last equality. We find that for $\chi >0$, the binary becomes more eccentric after the encounter with the DM particle, and for $\chi < 0$, the binary becomes more circular. 
Since the energy and angular momentum in the simulation are conserved to machine precision, we expect that the change in energy and angular momentum of the binary are equal and opposite to the change in energy and angular momentum of the DM particle. Therefore,
\begin{gather}
    \Delta E_{DM} = -\Delta E \\
    \Delta L_{z,DM} = - \Delta L_z
\end{gather}

For the sake of clarity, we use the normalized change in energy $\Delta \tilde{E}$ (as defined previously), change in angular momentum $\Delta \tilde{L_z}$, and eccentrification parameter $\tilde{\chi}$ in our calculations which are defined as follows: 

\begin{gather}
    \Delta \tilde{L_z} = \frac{\Delta L_{z,DM}}{L_{z,DM}^{c}} \\
    \tilde{\chi} = \chi \frac{M_2}{m_{DM}}
\end{gather}
where $L_{z,DM}^{c}$ is the angular momentum of a DM particle on a circular orbit with $a_{DM}=a_0$, and $E_{DM}$ is the binding energy of the DM particle.

To understand how the retrograde and prograde families contribute differently to the change in energy and angular momentum of the binary, we plot the distribution of $\Delta \tilde{E}$ and $\Delta \tilde{L_z}$ for a binary with $e=0.1,0.7$ in Figure \ref{fig:three_energy_mom_hist}. Quite surprisingly, we find that the energy exchanged during the scattering process is similar between the retrograde and prograde models with relative differences of at most $1-2$ percent between them. This is in tension with the DF theory where we would expect the retrograde family to have a much lower final energy than the prograde family owing to the larger relative velocity in the former case. $\Delta \tilde{E}$ is always positive signaling that \textit{all} particles contribute to dissipating energy from the binary, even fast moving particles with velocities larger than the velocity of the secondary, in contrast with the assumptions made in previous studies \citep[e.g.,][]{Kavanagh2020PhRvD.102h3006K,Becker2022PhRvD.105f3029B}.  Additionally, we note that the distribution of energy exchanged does not change substantially as the eccentricity is increased. Although not presented here, we find that the mean energy exchanged $\Delta \tilde{E}_{\mathrm{mean}}$ remains almost constant across all values of $e$, consistent with the findings from \citet{Sesana2011MNRAS.415L..35S}.

Contrary to the distribution of $\Delta \tilde{E}$, we find major differences between the retrograde and prograde models while comparing the distribution of $\Delta \tilde{L_z}$. We find that the distribution of angular momentum exchanged always skews towards lower values for prograde models compared to the retrograde models. The mean angular momentum exchanged is almost $37\%$ lower for the prograde family than the retrograde family when $e=0.1$. As the eccentricity increases, the differences become even more drastic. At $e=0.7$, the mean angular momentum exchanged by the retgrograde family is almost $10\times$ larger than the prograde family. Thus, we find that counter-rotating particles are more efficient at stealing angular momentum from the binary than co-rotating particles, especially for more eccentric binaries.

\begin{figure}
    \begin{center}
    \includegraphics[width=0.45\textwidth]{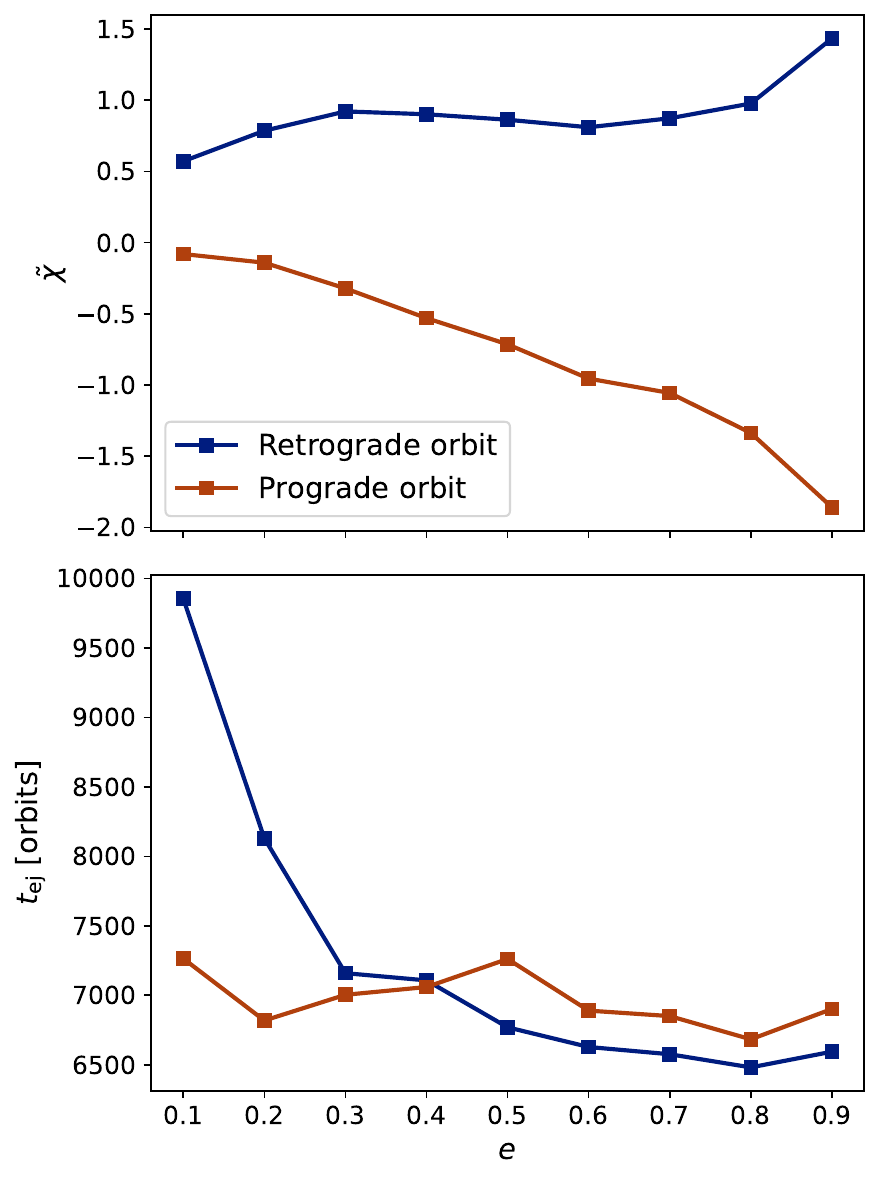}
    \caption{Top: the normalized eccentrification parameter $\tilde{\chi}$ as a function of the eccentricity $e$ of the binary. Bottom: the mean time of ejection of the DM particle $t_{\mathrm{ej}}$ in units of the number of orbits of the binary as function of the eccentricity of the binary. We notice that for retrograde models $\tilde{\chi}$ is always positive  and always negative for prograde models. This results in an eccentrification of the binary in case of retrograde rotation and circularization of the binary in case of prograde rotation. We also notice that the time of ejection of the DM particle is almost uniform for the prograde case across different values of $e$. For retrograde scenario, for larger values of $e$, $t_{\mathrm{ej}}$ is similar to that in the prograde case. However, it rapidly increases as $e$ decreases.   }
    \label{fig:three_body_chi}
     \end{center}
\end{figure}

We calculate the mean values of $\Delta \tilde{E}$ and $\Delta \tilde{L_z}$ across different binary eccentricities to calculate the eccentrification parameter using equation \ref{eq:sesana02}. We present the normalized eccentrification parameter $\tilde{\chi}$ as a function of $e$ in Figure \ref{fig:three_body_chi} for both the retrograde and prograde models. Consistent with \citet{Sesana2011MNRAS.415L..35S}, we find that $\tilde{\chi}$ is \textit{always positive} in retograde models while it is \textit{always negative} in prograde models. Thus, in our retrograde models, the binary always undergoes eccentrification as a result of interaction with the spike, whereas the binary undergoes circularization in the prograde scenario. This is consistent with findings from Fokker-Planck models of binary evolution in rotating nuclei. According to \citet{Rasskazov2017ApJ...837..135R} equation 84a,

\begin{equation}
    \langle \Delta e \rangle = \frac{\rho G a K H}{\sigma}
\end{equation}
 where $\langle \Delta e \rangle$ is the mean change in eccentricity, $H$ is the dimensionless binary hardening rate, and $K$ is the dimensionless eccentricity growth rate \citep[][]{Quinlan1996NewA....1...35Q,Sesana2006ApJ...651..392S}. Using Fokker-Planck methods, \citet{Rasskazov2017ApJ...837..135R} find that, in rotating nuclei,
 \begin{equation}
     K = 1.5 e (1-e^2) \left ( 0.15 - (2\eta-1) \rm{cos}(\theta) \right )
 \end{equation}
where $\theta$ is the orbital inclination of the binary and $\eta$ is the fraction of prograde to retrograde particles (as opposed to $\mathcal{F}$ which is the ratio of retrograde to prograde particles). Since our binary inclination is 0, $\rm{cos}(\theta)=1$. For a retrograde rotation, $\eta=0$ and we find, $K > 0$ so the binary eccentrifies as a result of interactions with DM particles. In the prograde case, $\eta=1$ and $K < 0$ and the binary circularizes due to three body interactions.
This explains why the binary eccentrifies or circularizes quickly at the beginning of our full $N$-body simulations. The rate of circularization is lower in the retrograde models than in the non-rotating and the prograde models, resulting in a faster inspiral as the effects of GW radiation are stronger at larger eccentricities. The inspiral is the slowest in prograde models as the binary circularizes faster than in non-rotating and even vacuum scenarios. While not presented here, we verified that our results are valid across the mass-ratios considered in this study. 

\begin{figure}
    \begin{center}
    \includegraphics[width=0.45\textwidth]{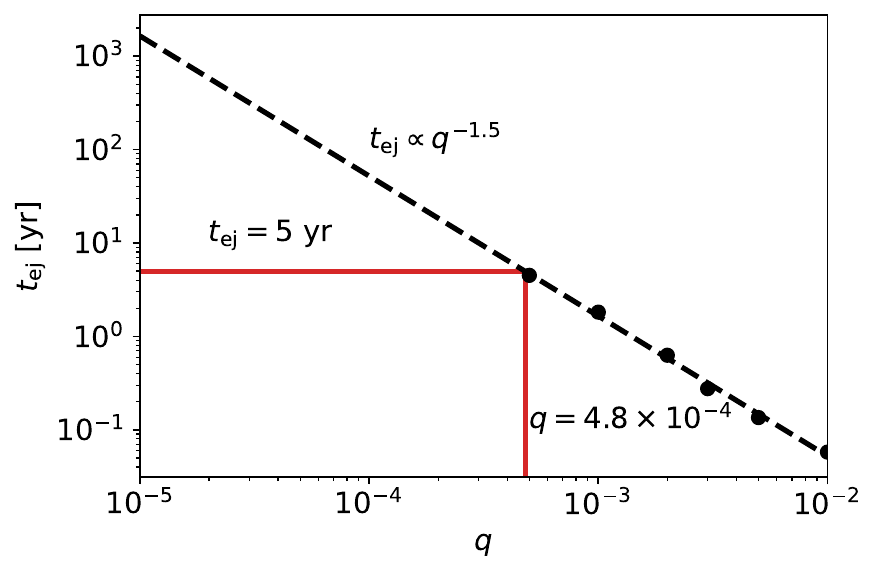}
    \caption{The mean ejection time $t_{\mathrm{ej}}$ as a function of the binary mass ratio $q$. The dots represent the  values calculated from the three-body simulations whereas the dashed line represents the best fit to the datapoints. We find that $t_{\mathrm{ej}} \propto q^{-1.5}$. This implies that for any $q < 4.8 \times 10^{-4}$, $t_{\mathrm{ej}} > 5$ yr, larger than the inspiral time of a LISA detectable IMRI. For $q=10^{-4}$ or lower, the ejection time is much larger implying that the backreaction on to the halo is expected to be minimal. This is consistent with the findings from our full $N$-body simulations for $q=10^{-4}$ where we did not find any substantial feedback on the DM density profile. } 
    \label{fig:tej_fitted}
     \end{center}
\end{figure}

One can also use the three-body simulations to understand how long it takes for the three-body interactions to disrupt the spike significantly. For a fixed mass ratio of $q=10^{-2}$, looking at Figure \ref{fig:three_body_chi}, we find that for lower eccentricity binaries, the mean ejection time $t_{\rm{ej}}$ is larger for particles on retrograde orbits than prograde orbits. As the eccentricity is increased, the ejection time becomes similar for both prograde and retrograde orbits. Since the particles are preferentially ejected when they are co-rotating with the binary, we postulate that difference between the prograde and the retrograde models across eccentricity is caused due to the fact that a less eccentric binary exterts weaker torques on the particle leading to a longer secular timescale over which the binary converts the particle from retrograde to prograde. For moderately or highly eccentric binaries, the conversion from retrograde to prograde rotation happens quickly, leading to a similar ejection time between the two models. 

We can also study the ejection time as a function of the mass ratio. We take a non-rotating version of the model generated above and run multiple simulations with different mass ratio binaries with $a_0=2\times10^{-8} \rm{pc}$, $e_0=0.7$ and plot the mean ejection time $t_{\mathrm{ej}}$ as a function of the mass ratio $q$ in Figure \ref{fig:tej_fitted}. We find that the relationship between the ejection time and the mass-ratio can be described by a power-law. We find that $t_{\mathrm{ej}} \propto q^{-1.5}$. The proportionality constant is a function of the central IMBH mass, semi-major axis, and eccentricity of the binary.

As pointed in section 3.2, we caution the reader, that our models are somewhat unphysical as $\mathcal{F}$, the fraction of counter-rotating particles to the total number of particles, is 0 (prograde models) or 1 (retrograde models). The eccentrification or circularization sensitively depends on this fraction $\mathcal{F}$. \citet{Sesana2011MNRAS.415L..35S} find that when this fraction is greater than 0.5, the binary eccentrifies as a result of three-body interactions. This suggests that in our non-rotating models, the binary undergoes mild eccentrification as a result of interactions with the spike. A systematic study of the change in the rate of circularization as a function of $\mathcal{F}$ is left for a future study. 

The three-body simulations are easier to parallelize, and faster to run than the $N$-body simulations which can provide the foundation for a more extensive parameter space study in the future. We can also use the three-body simulations to derive the distribution of the energy and angular momentum changes as a function of the binary semi-major axis and eccentricity which can later be used in semi-analytic models like {\tt\string HaloFeedback}. Future work might also involve considering the binary's angular momentum to be tilted relative to the angular momentum of the DM spike.

\subsection{Precession effects}

\begin{table}
\centering
\begin{tabular}{ll}
\hline
Model               & $|\Delta N^{(2)}|$ \\
\hline
Non rotating        & $2.4\times 10^5$       \\
Retrograde rotation & $2.6\times 10^5$       \\
Prograde rotation   & $1.9\times 10^5$      \\
\hline
\end{tabular}%
\caption{An estimate of the dephasing in the second harmonic $|\Delta N^{(2)}|$ of the GW signal for $q=10^{-3}$ models in $\gamma_{\rm{sp}}=7/3$ spike when relativistic precession effects are included. We find that the net dephasing for the non-rotating model is consistent among the non-precessing simulations and the precessing simulations. However, the effects of rotation are significantly dampened upon the inclusion of rotation leading to lower estimates of dephasing than before. }
\label{tab:precession_dephasing}
\end{table}

\begin{figure}
    \begin{center}
    \includegraphics[width=0.45\textwidth]{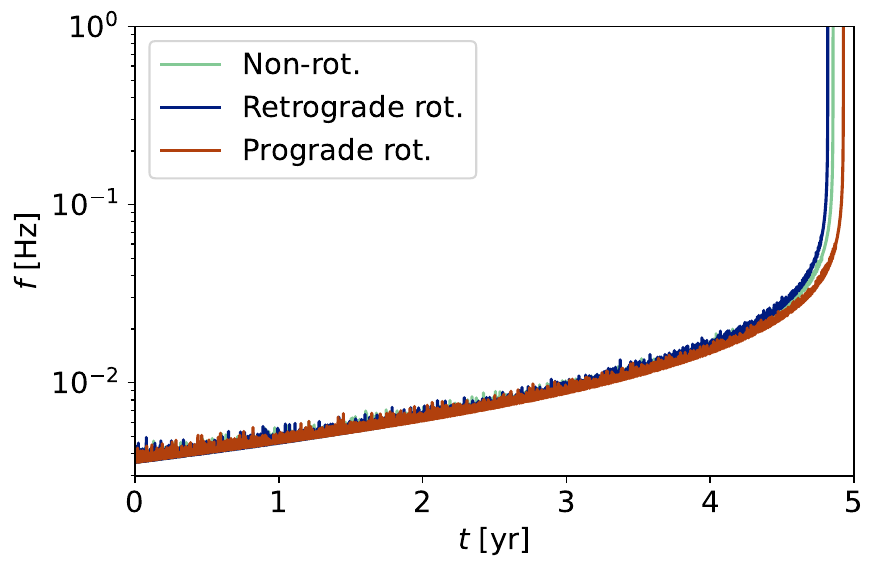}
    \caption{Evolution of the mean orbital frequency of the binary $f$ in Hz as a function of time $t$ in years for $q=10^{-3}$ in a $\gamma_{\rm{sp}}=7/3$ spike. We find that precession dampens the effects of rotating spikes substantially. In the non-precessing scenario the difference in dephasing between the non-rotating and rotating models was $\mathcal{O}(10^5)$ GW cycles, whereas in the precessing scenario that difference drops to $\mathcal{O}(10^4)$ GW cycles. Nevertheless, we find that the prograde rotation model takes longer to merge than the non-rotating model while the retrograde rotation model merges faster. This indicates that the results of our rotating models are robust, at least qualitatively.} 
    \label{fig:precession_evol}
     \end{center}
\end{figure}

We present a brief analysis of the effects of relativistic precession on the results from both our rotating and non-rotating models. Relativistic precession is included by using the PN1 and PN2 terms in a set of rotating and non rotating $q=10^{-3}$ models in $\gamma_{\mathrm{sp}}=7/3$ spikes. We plot the mean orbital frequency of the binary $f$ as a function of the inspiral time $t$ in Figure \ref{fig:precession_evol} for the rotating and the non rotating models. Comparing the evolution of the non rotating model in Figure \ref{fig:precession_evol} to that in Figure \ref{fig:evolution_non_rot}, we find that there are minimal differences in the evolution with and without relativistic precession.  On the other hand, we immediately notice that the inclusion of precession severely dampens the effect of the rotational effects that were evident in the non precessing simulations. Although, we find the same qualitative effects as in the non precessing simulations, i.e., retrograde model merges faster than the prograde and non rotating model, the results indicate that precession reduces the effects of rotation. Whereas in the non-precessing retrograde model, the binary takes 433.4 fewer days to merge compared to the inspiral in vacuum, in the precessing retrograde model, the binary takes 145.4 fewer days to merge. In the non-precessing prograde models, we found that the binary actually took 227.8 days longer to merge than in vacuum but we find that upon the inclusion of precession it merges 105 days earlier than in vacuum. We note that this is still $\sim$ 38 days slower than the non rotating model. The changes in the inspiral time are also reflected in the amount of dephasing over the 5 year inspral timespan. We present the number of dephasing cycles of the second harmonic $|\Delta N^{(2)}|$ over the course of the full inspiral in Table \ref{tab:precession_dephasing}. The amount of dephasing for the non rotating model is consistent among both sets of simulations, i.e. precessing and non precessing, but we notice differences in the rotating models. In the non-precessing retrograde model $|\Delta N^{(2)}| \approx 7 \times 10^5$ but in the precessing retrograde model that drops to $2.4 \times 10^5$. We note that in the non precessing prograde models, we required $5\times10^5$ more cycles compared to vacuum for the binary to merge but due to the decreased effect of rotation, the binary merges faster, taking about $1.9\times10^5$ fewer cycles. The differences in dephasing between the non rotating and rotating models in the non-precessing case was $\mathcal{O}(10^5)$ but upon the inclusion of relativistic precession, that difference drops to $\mathcal{O}(10^4)$. 

The reason for the drastic changes in the rotating models upon the inclusion of precession is unclear. \citet{Sesana2011MNRAS.415L..35S} noted that in self-consistent $N$-body simulations where Newtonian precession of an extended gravitational system played a role, the effects of rotation were dampened. Here we observe a similar effect but with relativistic precession indicating the similarities between the two scenarios. Notably, the effect of the eccentric binary on the secular evolution of a particle can be described through the lens of the eccentric Kozai-Lidov effect \citep[e.g.,][]{Merritt2009ApJ...693L..35M,Merritt2013degn.book.....M}. Recent simulations of hierarchical MBH triplets have found that inclusion of PN effects can diminish or even extinguish the Kozai-Lidov evolution of the inner binary \citep[][]{Tanikawa2011ApJ...728L..31T,Bonetti2018MNRAS.477.3910B,Mannerkoski2021ApJ...912L..20M,Romulus2023A&A...678A..11K}. These simulations may hold clues to understanding the effect of precession and how it affects our results. However, a more thorough analysis requires future work.

Although the effects of rotation are decreased in the $q=10^{-3}$ models, we expect rotation to still play a significant role in lower mass ratio binaries where the mass enclosed within the orbit of the binary would be larger. In such a case, the Newtonian precession of the spike could also play a part and the net change in precession could be used as an indicator for the presence of DM. However verifying this requires simulations that are beyond the scope of this current study.

\section{Discussion} \label{sec:discussion}
\subsection{Accretion effects}
The secondary is also expected to accrete from the spike during the inspiral. For the parameters considered in our simulations, the accretion effect is expected to be quite minimal, especially when $M_2 = 1 M_{\odot}$. For the stellar mass BH scenario with $M_2 = 10 M_{\odot}$, we can estimate the rate of accretion assuming that the secondary undergoes Bondi-Hoyle accretion \citep{Bondi1944MNRAS.104..273B,Edgar2004NewAR..48..843E,Macedo2013ApJ...774...48M,machPhysRevLett.126.101104}. Following \citet{Yue2018PhRvD..97f4003Y}, the change in mass of the secondary over time ($\frac{dM_2}{dt}$) can be written as
\begin{gather} \label{eq:accretion}
    \frac{dM_2}{dt} = \frac{16 \pi G^2 M_2^2 \rho_{\mathrm{DM}}}{c^2 v} \left ( 1+\frac{v^2}{c^2} \right )
\end{gather}
where $v$ is the velocity of the secondary. Assuming that $\rho_{\mathrm{DM}} \sim 10^{20} M_{\odot} \mathrm{pc}^{-3}$ near the secondary, and $v \sim 2\times10^4 \mathrm{km} \mathrm{s^{-1}}$, we find that $\frac{dM_2}{dt} \sim 0.005 M_{\odot} \mathrm{yr}^{-1}$. Since the spike gets disrupted within the first 0.1 yr, we expect the accretion onto the secondary to be of the order of $10^{-4} M_{\odot}$ having minimal effects on our results. One can imagine that in lower mass ratio scenarios where the spike is not disrupted as much, the accretion effects might be larger, but still subdominant compared to dephasing due to three-body scattering.

We  note that $\frac{dM_{2}}{dt} \propto \frac{1}{v}$. Since the relative velocity between the DM particles is lower in prograde rotating models than isotropic or retrograde models, the accretion effects could be larger. An estimate of the difference is beyond the scope of this current study but will be considered in the future.

\citet{Nichols2023arXiv230906498N} recently explored a self-consistent treatment of accretion in IMRIs with stellar mass BHs as the secondary and found that inclusion of accretion can lead to difference of $100-1000$ GW cycles compared to the models where accretion is not included. Although this represents about $<1$ per cent difference in the number of dephasing cycles in our $q=10^{-3}$ and $10^{-4}$ models, future studies will need to account for accretion to generate LISA waveforms since matched filtering requires a proper waveform determination over a few cycles. More importantly, the change in eccentricity due to accretion would also need to be studied considering the large effect eccentricity plays in GW radiation dominated binary evolution, as demonstrated in this work.


\subsection{Dark matter annihilation and EM signatures}

In addition to modifying GW signals of IMRIs, DM spikes are considered to be a source of gamma radiation as a result of DM annihilation. Thus, they provide strong indirect signatures of weakly interacting DM particles and allow us to probe DM microphysics \citep[e.g.,][]{Gondolo1999PhRvL..83.1719G,Ullio2001PhRvD..64d3504U,Fields2014PhRvL.113o1302F,Shapiro2016PhRvD..93l3510S,Lacroix2018A&A...619A..46L}. In our work we do not consider the effect of DM annihilation on the spike profile. Our models are, therefore, representative of the non-annihilating DM case. Nevertheless, we provide estimates on how the spike profile can be changed and discuss how that would affect our results in case of annihilating DM. 

For self-annihilating DM, the spike profile is significantly depleted due to the interactions between DM particles.  It has been suggested that near the central MBH, a flat plateau, or core forms as a result of DM annihilation \citep[][but see also \citet{Vasiliev2007PhRvD..76j3532V,Shapiro2016PhRvD..93l3510S}]{Gondolo1999PhRvL..83.1719G}. The density of this plateau $\rho_{\mathrm{pl}}$ is given as 

\begin{gather}
    \rho_{\mathrm{pl}} = \frac{m_{\chi}}{\langle \sigma v \rangle T}
\end{gather}
where $m_{\chi}$ is the mass of the DM particle, $\langle \sigma v \rangle$ is the interaction cross section, and $T$ represents the age of the MBH. The interaction cross-section of the DM particles is considered to be a constant in standard thermal weakly interacting particle (WIMP) models. Assuming a scenario with $m_{\chi} = 1 \mathrm{TeV}$, and $T \gtrsim 10^6 {\mathrm{yr}}$ we can calculate the density of the plateau and the radius of the plateau ($r_{\mathrm{pl}}$) by setting $\rho_{\mathrm{DM}}(r_{\mathrm{pl}})=\rho_{\mathrm{pl}}(r_{\mathrm{pl}})$ for various values of $\langle \sigma v \rangle$. We present this information in Table \ref{tab:annihilation}. We find that unless the DM annihilation cross section is very small, the density of the plateau is lower than the density of the spike where the binary is situated. The lower density would lead to a smaller amount of dephasing. Since the dephasing is proportional to the density of DM near the binary, as argued before, when $\langle \sigma v \rangle=10^{-27}$ and $10^{-30} \mathrm{cm}^3 \mathrm{s}^{-1}$, the dephasing would be reduced by factors of $50-5\times10^5$. In such a case, the dephasing would be significantly reduced, and non-detectable in the $\langle \sigma v \rangle=10^{-27} \mathrm{cm}^3 \mathrm{s}^{-1}$ case. However, it should be noted that, according to \citet{Magic2016JCAP...02..039M} the upper limit on the cross-section is $\sim 10^{-25} \mathrm{cm}^3 \mathrm{s}^{-1}$ and the annihilation cross-section in reality could be much lower. In such a scenario, one could observe GW dephasing in tandem with an EM signature.

\begin{table}
\centering
\begin{tabular}{llll}
\hline
$M_1 [M_{\odot}]$                                                & $\langle \sigma v \rangle [\mathrm{cm}^3 \mathrm{s}^{-1}]$ & $\rho_{\mathrm{pl}} [M_{\odot} \mathrm{pc}^{-3}]$                                         & $r_{\mathrm{pl}} [\mathrm{pc}]$                                             \\
\hline

\begin{tabular}[c]{@{}l@{}}$10^3$\\ $10^4$\end{tabular} & $10^{-27}$   & $8.4\times10^{14}$ & \begin{tabular}[c]{@{}l@{}}$2.2\times10^{-6}$\\ $4.8\times10^{-6}$\end{tabular} \\
\hline

\begin{tabular}[c]{@{}l@{}}$10^3$\\ $10^4$\end{tabular} & $10^{-30}$   & $8.4\times10^{17}$ & \begin{tabular}[c]{@{}l@{}}$1.1\times10^{-7}$\\ $2.5\times10^{-7}$\end{tabular} \\
\hline

\begin{tabular}[c]{@{}l@{}}$10^3$\\ $10^4$\end{tabular} & $10^{-33}$   & $8.4\times10^{20}$& \begin{tabular}[c]{@{}l@{}}$5.9\times10^{-9}$\\ $1.3\times10^{-8}$\end{tabular} \\ 
\hline

\end{tabular}%
\caption{The annihilation radius $r_{\mathrm{pl}}$ and annihilation plateau density $\rho_{\mathrm{pl}}$ under different DM cross sections $\langle \sigma v \rangle$ for a $\gamma_{\mathrm{sp}}=7/3$ spike with a central BH with mass $M_1$. }
\label{tab:annihilation}
\end{table}

We note that the above mentioned case is an optimistic scenario. Under a different scenario, we consider $T \sim 10^{10} \mathrm{yr}$, 
typical for the ages of nearby galaxies. Using more conservative values of $m_{\chi} = 35 \mathrm{GeV}$ and $T=10^{10} \mathrm{yr}$ leads to a plateau density of $2.9 \times 10^9 M_{\odot} \mathrm{pc}^{-3}$ (assuming that $\langle \sigma v \rangle = 10^{-27} \mathrm{cm}^3 \mathrm{s}^{-1}$) which would leave no imprints on the GW signal. It should be noted, however, that over such a long duration, the system is not expected to be isolated. 

Alternately, one could use the detection of GW signals from DM spikes to put upper-limits on the annihilation cross section of DM. Setting $\rho_{\mathrm{pl}}=\rho_{\mathrm{DM}}$, we get
\begin{gather}
    \langle \sigma v \rangle \leq \frac{m_{\mathrm{\chi}}}{\rho_{\mathrm{sp}} (\frac{r_{\mathrm{sp}}}{r_{\mathrm{pl}}})^{\gamma_{\mathrm{sp}}} T}
\end{gather}
where we used equation 7 for $\rho_{\mathrm{DM}}$. For $M_1=10^3 M_{\odot}$ with a $\gamma_{\mathrm{sp}}=7/3$ spike, $r_{\mathrm{sp}} \approx 0.5$ pc. As in our previous approximations, we take $\rho_{\mathrm{sp}}=226 M_{\odot} \mathrm{pc}^{-3}$. We assume that the annihilation radius $r_{\mathrm{pl}} \sim a = 2\times10^{-8}$ pc, the semi-major axis of the binary. Since the dephasing due to three-body scattering depends on the local density as mentioned in previous sections, we should expect to obtain a similar amount of dephasing in the annihilation scenario as in the non-annihilating scenario. We, then, obtain the following upper-limit on the cross section:
\begin{equation}
    \langle \sigma v \rangle \lesssim 1.71 \times 10^{-32} \mathrm{cm}^3 \mathrm{s}^{-1} \left ( \frac{m_{\chi}}{1 \mathrm{TeV}} \right ) \left ( \frac{10^6 \mathrm{yr}}{T}\right).
\end{equation}
Similar analysis can be performed with different spike parameters and central IMBH masses. We point to the reader, however, that a lower amount of dephasing than expected can arise from different DM properties that lead to a larger annhilation, or from dynamical factors such as rotation, as pointed in this study. In such a scenario, a different signature, possibly electromagnetic would be needed to break the degeneracy. In any case, as \citet{Hannuksela2020PhRvD.102j3022H} points out, any detection of DM spike using GWs will be in strong tension with current thermal WIMP models and place  constraints on the mass of the particle.

Other models can suggest a lower cross section allowing for the co-existence of EM signals along with GW signals from the spike \citep[e.g.,][]{Shelton2015PhRvL.115w1302S}. However, \citet{Hannuksela2020PhRvD.102j3022H} report that in such a case, even in the most optimistic scenario with $M_1=10^6 M_{\odot}$, the electromagnetic counterpart will be detectable only up to a distance of 90 Mpc by detectors such as ASTROGAM \citep{Astrogam2018JHEAp..19....1D}, FERMI, or CTA. Only a few IMRI events would happen in such a small volume and an EM counterpart would require a large fraction of IMRIs to be embedded in DM spikes. Since the luminosity of the gamma radiation $L$ is proportional to the squared density of the DM spike profile $\rho_{\mathrm{DM}}^2$, the EM signals from IMBHs are expected to be weaker and the prospects for finding EM counterparts are more pessimistic.   

\subsection{Implications for binary inspiral in realistic environments} \label{subsec:realistic_spike}
As we have seen in the previous sections, the amount of dephasing is  sensitive to the density profile near the IMBH. Furthermore, the actions of the inspiraling compact object can have a drastic effect on the spike, completely unbinding it. This implores us to ask an important question: how realistic are the spikes considered in this study and previous studies? How do realistic spikes affect the dephasing of an inspiraling compact object? While a comprehensive study to understand spike profile in realistic environments is beyond the scope of this work, we present a brief analysis of the impact of surrounding environments and past inspirals on the spike density profile. 

In a non-isolated environment interactions between the spike and the surrounding material can reduce the density of the spike compared to isolated adiabatic growth models. Previous works have only considered the effect of stars surrounding the spike \citep[e.g.,][]{Merritt2007PhRvD..75d3517M}. Using the Fokker-Planck code \texttt{Phaseflow} \citep{Vasilev2017ApJ...848...10V} we study, for the first time, the effect of stellar mass BHs along with stars on the final DM spike profile. Here we only consider the case where the central SMBH has a mass of $10^6 M_{\odot}$ and the total mass of the stellar mass BH particles is 1\% of that of the total mass of stars in the galactic nucleus. The total stellar mass in the nucleus is set to be $10^7 M_{\odot}$. We note that this excludes the bulge mass. Our two-component model is consistent with the expectation from the Kroupa initial mass function (IMF) \citep{Kroupa2001MNRAS.322..231K} The stellar mass BH particles are given masses  10 times larger than that of the star particles. The DM halo has a Hernquist profile initially representing a $\gamma=1$ slope in the center and the stars and BHs in the nuclei are given a shallow $\gamma=0.5$ cusp.

\begin{figure}
    \begin{center}
    \includegraphics[width=0.45\textwidth]{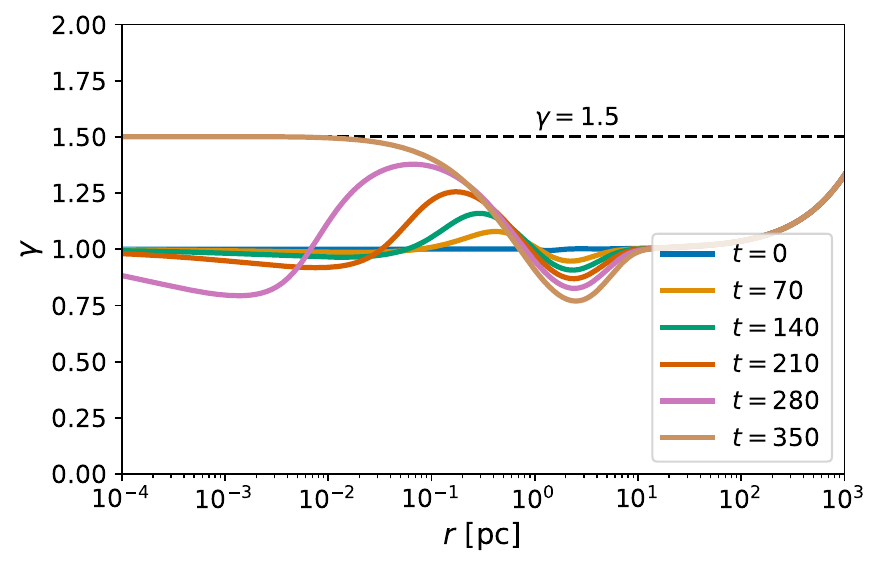}
    \caption{Slope of the DM profile $\gamma$ as a function of distance $r$ from the central MBH at various times. The time is presented in $N$-body units with 1 time unit corresponding to $\approx 14.9$ Myr. The initial profile of the DM surrounding the MBH is a $\gamma=1$ model, representing a Hernquist type halo. The DM profile is embedded in a nuclei consisting of stars and stellar mass BHs, drawn from a Kroupa IMF, and the subsequent evolution is performed using a Fokker-Planck model. We find that, similar to the stars only model, the spike reaches a $\gamma=1.5$ profile at the center, near the MBH. However, unlike the stars only model, two body relaxation is enhanced due to the presence of a two component mass function, leading to an accelerated growth.  } 
    \label{fig:fokker_planck}
     \end{center}
\end{figure}

In Figure \ref{fig:fokker_planck}, we plot the slope of the DM component at different times and find that in an equilibrium state the spike reaches a $\gamma = 1.5$ profile near the MBH. This is similar to what is observed in a stars only model. However, due to mass segregation in the presence of a two component mass spectrum, the rate of growth of the spike is enhanced. Comparing our two component mass model to that from a single component model, we find that presence of two mass species accelerates the growth by a factor of $4$. In fact, this is quite sensitive to the IMF and the initial density profile of the nuclei. A top-heavy IMF \citep[e.g.,][]{Chabrier2005ASSL..327...41C} results in an even faster growth. This can have major implications for the time taken by the spike to regrow after it has been disrupted. A detailed study of regrowth time after disruption will be considered in a future work.

A major caveat of the above result is that we do not consider the effect of stellar evolution on relaxation. Stellar evolution can lead to mass loss and kicks to compact objects over time which reduces their population and therefore affects the relaxation timescale. We consider the effect of stellar evolution in an approximate manner by evolving a population of stars drawn from a Kroupa IMF with different metallicities and in different environments using SSE \citep{Hurley2000MNRAS.315..543H}. In a dense environment such as a galactic nucleus where the escape velocity is higher, we find the mass-ratio of BH particles approaches  $M_{\rm BH}/M_{\rm *} \sim 10^{-3}$ of the mass of stellar particles after 5 Gyr of evolution when the metallicity $Z=0.1 Z_{\odot}$ where $Z_{\odot}$ is the solar metallicity.  This drops to  $M_{\rm BH}/M_{\rm *} \sim 10^{-4}$ when $Z= Z_{\odot}$. The decrease in mass is due to stellar evolution mass loss and random kicks imparted by SSE onto compact objects during their formation. In a globular cluster like environment with a total mass of $10^6 M_{\odot}$, the escape velocity is lower and we find heavier compact objects are hardly retained ($M_{\rm BH}/M_{\rm *} \sim 0$). This affects the timescale of growth of the DM spike. Using Fokker-Planck models with the evolved profile, we find that when $M_{\rm BH}/M_{\rm *} \sim 10^{-3}$ , the timescale of spike growth increases by 40\% compared to the two-component model without stellar evolution where $M_{\rm BH}/M_{\rm *} \sim 10^{-2}$. When $M_{\rm BH}/M_{\rm *} \sim 10^{-4}$, the timescales increases by a factor of $2$  compared to the non-stellar evolved model. This highlights the importance of inclusion of a mass-species which accelerates the growth of a spike even when the population of the heavier mass species is much smaller than lower mass species.

Once a spike has been disrupted, the regrowth happens on timescales that are on the order of collisional relaxation time within the sphere of influence of the MBH. The relaxation time is affected by the mass species surrounding the MBH. In a single component model, relaxation takes longer than when a mass spectrum is present, as evident from our results above. We can estimate the relaxation time $t_{\rm relax}$ in a stellar only environment as follows \citep[e.g.,][]{Babak2017PhRvD..95j3012B,Becker2024arXiv240402808B} :

\begin{equation}
    t_{\rm relax} = \frac{5}{\rm{ln}(\Lambda)} \left ( \frac{\sigma}{10 \rm{km} \rm{s}^{-1}} \right ) \left ( \frac{r_{\rm infl}}{1 \rm{pc}} \right )^2 \rm {Gyr}
\end{equation}
where $\rm{ln}(\Lambda)$ is the Coulomb logarithm,$\sigma$ is the velocity dispersion, and $r_{\rm infl}$ is the influence radius of the MBH. $\sigma$ can be estimated from the well known $M-\sigma$ relationship \citep[e.g.,][]{Guletkin2009ApJ...698..198G, Kormendy2013ARA&A..51..511K} or from density profiles of galactic nuclei. Since we focus on IMBHs in this work, we refer to the density profiles of known dwarfs from \citet{Nguyen2017ApJ...836..237N,Nguyen2018ApJ...858..118N} to calculate $\sigma$ and $r_{\rm infl}$. Taking $\rm{ln}(\Lambda) \sim 10$, $\sigma \sim 50 \rm{km} \rm{s}^{-1}$, and $r_{\rm infl} \sim 0.1$ pc ,we find $t_{\rm relax} \lesssim 0.025$ Gyr for a $10^4 M_{\odot}$ IMBH. This is representative of relaxation time in a high density environment. On the other hand, in a low density environment, $r_{\rm infl}$ is going to be larger. Assuming $r_{\rm infl} \sim 0.5$ pc in that case, we find $t_{\rm relax} \lesssim 0.6$ Gyr. The inspirals happen on timescales over which two-body relaxation depletes the cusp by driving the compact objects on loss cone orbits. The depletion timescale $t_{\rm d}$ is estimated as follows \citep[e.g.,][]{Babak2017PhRvD..95j3012B,Becker2024arXiv240402808B}:
\begin{equation}
    t_{\rm d} = \frac{20}{1 + R} \left ( \frac{m}{10 M_{\odot}} \right)^{-1} \left( \frac{M_{\rm{MBH}}}{10^6 M_{\odot}} \right)^{1.19} \rm {Gyr}
\end{equation}
where $R$ is the ratio of plunges to inspirals, and $m$ is the characteristic mass of the compact object. Taking $R=0$ to find the upper limit and $m=10 M_{\odot}$ representing BHs , we find that $t_d \lesssim 0.1$ Gyr. Thus, the time between inspirals is comparable to the two body relaxation time in high density environments and is shorter in the low density environments. As such, we  expect the spike to not exist in its equilibrium state in low density environments. However, under the presence of a mass spectrum the relaxation time decreases in which case $t_d \sim t_{\rm relax}$ which can lead to the spike existing in the equilibrium $\gamma=1.5$ state, even in low density environments. This highlights the importance of examining the spike growth embedded in a realistic mass spectrum. We note one caveat of this calculation is that relaxation times after depletion of a spike a larger than the ones calculated from the above equation. Therefore, our relaxation timescale should be considered as a lower limit. 

\begin{figure}
    \begin{center}
    \includegraphics[width=0.45\textwidth]{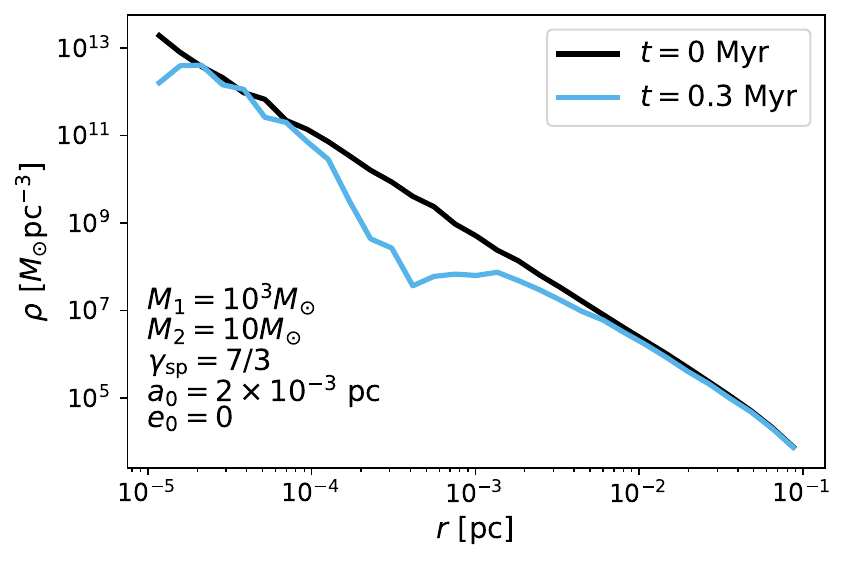}
    \caption{Density of a pristine $\gamma_{\rm sp} = 7/3$ DM spike $\rho$ as a $10 M_{\odot}$ BH is in. Even at larger separations, we find that the inspiraling object forms a core extremely rapidly. This suggests that dense spikes of the form of $\gamma_{\rm sp} = 7/3$ can only be present if there were no prior inspirals. }
    \label{fig:realistic_core_form}
     \end{center}
\end{figure}

A big question also lingers regarding the starting point of the compact object in the simulations. In a realistic scenario, a compact object would be unbound initially and become bound over time due to the effect of dynamical friction, two-body relaxation, and hardening in the stellar and DM environment. In a stellar dominated environment the compact object is going to be driven to inspiral orbits mainly due to the effects of two-body relaxation. In such a case, we would expect the initial inspiral to cause minimal disruption to the spike. However, in such a case, the spike profile would have a $\gamma_{\rm sp}=1.5$ slope which we showed produces minimal dephasing effects for the systems considered. This was also highlighted in \citet{Becker2024arXiv240402808B} where the author found that GW signals from the spikes in this case would be hardly detectable. On the other hand, in an isolated environment with no erosion, the compact object is going to inspiral due to the effect of dynamical friction and three body hardening from the DM environment. In such a case, the formation and inspiral itself would produce a flat core that can massively reduce dephasing effects. As an example, we produce the impact on the density profile from a simulation with a $\gamma_{\rm sp} = 7/3$ spike where the compact object now starts at an initial semi-major axis of $a_0=2\times10^{-3}$ pc in Figure \ref{fig:realistic_core_form}. Within 0.1 Myr, we find a core has formed with a density of $\sim 10^{7} M_{\odot} \rm{pc}^{-3}$. According to \citet{MerrittSzell2007ApJ...671...53M}, the formation of a hard binary is accompanied by an ejection of mass comparable to the mass of the binary. Therefore in DM only environments, we expect the density to be even lower and the core to be larger. The relaxation time of the spike in this isolated environment is so long that we do not expect a regrowth within a Hubble time. This indicates that the simulations with the $\gamma_{\rm sp}=7/3$ models represent optimistic scenarios where the spike \textit{exists in isolation and has not undergone a previous inspiral}. This is a major caveat of our work and that of previous works. Additionally, on account of their lower masses, IMBHs can be off-center where the adiabatic growth of the spike can be diminished. In such a scenario, a $\gamma=1.5$ is typically formed, as noted by \citet{ZhaoPhysRevLett.95.011301}. All of the above noted issues are pertinent and we emphasize the need to understand spike growth in realistic environments. We leave this as a future study.

\subsection{Prospects for GW detection}

As shown in Figure \ref{fig:strain_ic}, IMRI sources are promising for multiband GW astronomy with LISA and potential decihertz detectors. 
Several formation scenarios have been proposed for the formation of IMRIs in DM spikes, including host sites such as the nuclear star clusters of dwarf galaxies \citep{Yue2019ApJ...874...34Y} and merger remnants of elliptical galaxies \citep{Vasquez2023MNRAS.518.2113V}. 
The merger rate of such systems with and without DM spikes, however, is still uncertain.  
More detailed population synthesis of these sources with either semi-analytic models, cosmological simulations, or some combination thereof would provide more insight into what merger rates might be possible.
Conversely, the GW detection of IMRI sources and constraints on potential DM environments would provide tests of astrophysical population models in addition to DM physics. 

In light of the critical examination of realistic spike profiles from the previous section and \citet{Becker2024arXiv240402808B}, the prospects for dephasing due to DM spikes on IMRIs appear to be somewhat not optimistic. However, we emphasize that a full parameter space study is required before we can conclusively determine whether spikes would have tangible dephasing impact on GWs from IMRIs. On the other hand, \citet{Dai2022PhRvD.106f4003D,Dai2023arXiv230105088D} and \citet{Becker2024arXiv240402808B} noted that inspirals often happen on highly eccentric orbits with larger semi-major axes. In such a case, the DM spike can lead to a periapse precession in addition to the relativistic precession. The Newtonian precession of the spike is opposite in direction to the relativistic precession and the net effect can be quantified as a \textit{deshifting} index \citep{Becker2024arXiv240402808B}, which is a measure of the change in GW cycles due to change in pericenter precession from the DM spike. \citet{Dai2023arXiv230105088D} noted that deshifting can be present even in low density environments. Our preliminary analysis suggests that deshifting may be a more optimistic signature than dephasing. Unfortunately, for the set of models considered in this work, we were not able to quantify deshifting. In the future, we plan on studying  an extended parameter space where we study dephasing as well as deshifting across spike models.

\par

A potential space-based decihertz GW detector provides a significant advantage for boosting the IMRI detection rate as well as breaking parameter degeneracies that might be encountered with just LISA data. 
In addtion, if the inspiral phase of IMRIs is at a sub-threshold level in LISA data, the resolved merger phase in a space-based decihertz GW detector can provide priors to search for the subthreshold inspiral phase in archival LISA data. 
This is similar to how LIGO and 3G/XG ground-based detectors like Einstein Telescope or Cosmic Explorer may provide prior information on stellar-mass BH binary mergers to search for potential inspiral counterparts in archival LISA data \citep{Ewing2021PhRvD.103b3025E}.

In addition to the proposed tests of dynamical friction and accretion on gravitational waveforms, one can also test for the presence of three-body/loss-cone scattering for a potential IMRI event. 
One caveat for testing the presence of such effects might be that nonlinear feedback and DM-spike relaxation may complicate waveform parameterization. 
Though this certainly motivates more numerical simulations of IMRIs in DM spikes in order to explore the parameter space more comprehensively and to inform waveform parameterizations that are interpretable. 


\section{Conclusions} \label{sec:conclusion}
IMRIs are considered among the most crucial sources of low-frequency gravitational waves (GWs) that future space-based GW detectors like LISA and DECIGO can potentially detect. Recent studies have suggested that if the  intermediate-mass black hole (IMBH) in these systems is surrounded by a dark matter (DM) spike, the gravitational effects of the spike may induce dephasing in the observed GW signal, serving as a distinctive signature of the DM spike. Prior investigations have predominantly employed analytic methods to model the interaction between the spike and the binary, treating the gravitational impact of the spike on the binary through dynamical friction (DF). However, these approaches neglect the influence of the binary on the spike itself, a factor that can significantly affect the dynamics and, consequently, the degree of dephasing.

In our study, we employ $N$-body methods to delve into the complex dynamics involved in eccentric IMRIs embedded in both non-rotating and rotating DM spikes, evolving the system self-consistently. Our work represents the first attempt at modeling such systems using $N$-body simulations. The simulations reveal that, contrary to previous assumptions, the primary mode of dissipating energy from the system is not DF, a cumulative effect of two-body encounters, but rather three-body scattering, similar to stellar loss-cone scattering in SMBH/MBH binaries. We note that during the submission of this work, \citet{Kavanagh2024arXiv240213762K} also finalized their work on $N$-body simulations employing methods similar to ours. They qualitatively find similar behavior of the IMRI but find that the interactions can be described using a semi-analytic method combining dynamical friction loss with the time dependent potential of the binary. While their simulations are run for a shorter duration than ours, the combined effect which they propose is of three-body nature and similar to what we describe in this work. 

The main conclusions of our work can be summarized as follows:
\begin{itemize}
    \item The evolution of the non-rotating models reveals that in larger mass ratio cases, the binary is highly efficient at rapidly degrading the spike, resulting in minimal dephasing effects. Depending on the mass-ratio of the binary, our dense spike models with $\gamma_{\rm sp}=7/3, 9/4$ show that dephasing of $\mathcal{O}(10^4) - \mathcal{O}(10^5)$ can be expected.  However, in lower density $\gamma_{\rm sp} = 3/2$ spikes produced in a collisional environment, we find minimal dephasing of $\leq \mathcal{O}(10)$.
    \item We find that all of our simulations with dense spikes ($\gamma_{\rm sp}=7/3, 9/4$) predict that the dephasing is much larger, by factors of 10-100, than that predicted by the self-consistent semi-analytic method {\tt\string HaloFeedback} for similar binaries in circular orbits. This indicates that DF theory is unable to fully capture the dynamics of the binary.
    \item In our rotating models with dense $\gamma_{\rm sp}=7/3$ spikes, we find that, spikes that counter-rotate with the binary lead to faster inspirals compared to spikes that co-rotate with the binary, which lead to slower inspirals, even slower than inspirals in vacuum. This leads to dephasing effects that are $2.5-3.5$ times higher in the retrograde/counter-rotating cases than their non-rotating counterparts.
    \item We use three-body simulations to investigate the nature of the interaction of the binary with spike particles and find that the binary primarily dissipates energy via three-body interactions rather than DF, a two-body effect. The impact of DF in all of our simulations is sub-dominant shedding some light on the discrepancy between our results and those obtained in previous studies. Additionally, three-body simulations reveal that in rotating spikes, particles on retrograde motion eccentrify the binary whereas particles on prograde motion circularize the binary. This indicates that rotation in spikes cannot be neglected and should be the subject of further investigations in the future. 
    \item  We conduct simulations to investigate the impact of relativistic precession on dephasing in our $q=10^{-3}$ models. Our findings indicate that precession has minimal to no effect on dephasing in the non-rotating models but significantly diminishes the effects of rotation.
    \item By examining our initial models with Fokker-Planck methods, we assess the presence and detectability of spikes in realistic environments. Our results suggest that non-isolated environments have DM spikes with shallower slopes than previously considered, leading to smaller signals and lower detection prospects via dephasing.  We show that the densest DM spike models with $\gamma_{\rm sp} = 7/3$ represent scenarios where the spike is isolated and has not undergone any previous inspirals, as any inspiral leads to the formation of a low-density core where the regeneration of the spike takes much longer than the Hubble time. In the presence of lower density $\gamma_{\rm sp}=3/2$ spikes, present in non-isolated environments, dephasing is minimal and non-detectable. High stellar and stellar-mass black hole densities can accelerate the growth of $\gamma_{\rm sp}=3/2$ spikes, but a more extensive parameter space study is needed to determine which E/IMRIs embedded in such spikes can produce detectable dephasing signatures. Our preliminary analysis, coupled with recent studies, suggests that "deshifting" rather than dephasing might be a more optimistic signature, as it is more robust even in low-density environments \citep[][]{Dai2022PhRvD.106f4003D, Becker2024arXiv240402808B}.

\end{itemize}

Although our work uses idealized models, it provides a foundational exploration that paves the way for more comprehensive analyses in the future. Despite critical uncertainties, particularly regarding the growth, retention, and population of DM spikes around IMBHs, our research highlights the promise and potential benefits of delving deeper into this intricate problem. We demonstrate the need to consider more realistic environments, as detection prospects depend sensitively on spike density, which is influenced by the surrounding environment. Our work also underscores the importance of multiband GW astronomy and the need for a dedicated decihertz detector. By enhancing our approach to include accretion onto the inspiraling object and incorporating higher-order post-Newtonian terms, our $N$-body code emerges as an effective method for simulating the evolution of IMRIs embedded in DM spikes.

\section*{Acknowledgements}

 We thank the anonymous referee for invaluable comments on this work. We thank Tiziana Di Matteo and Katelyn Breivik for helpful feedback and discussions. We thank Qirong Zhu for helpful discussions regarding accurate integration schemes. We thank Niklas Becker for comments on this work. We thank Bradley Kavanagh for useful discussions and comments on this work.  We acknowledge the usage of the McWilliams center Vera computing cluster at the Pittsburgh Supercomputing Center. DM acknowledges support from the McWilliams Center-Pittsburgh Supercomputing Center seed grant and NASA grant 80NSSC22K0722. HT acknowledges support from NASA grants 80NSSC22K0722 and 80NSSC22K0821. GO was supported by the National Key Research and Development Program of China (Grant No. 2022YFA1602903), the National Natural Science Foundation of China (Grant No. 12373004) and the Fundamental Research Fund for Chinese Central Universities (Grant No. NZ2020021, No. 226-2022-00216).

\section*{Data Availability}

The data from the simulations is available upon reasonable requests. The code used in this work will be made publicly available upon the acceptance of this paper.



\bibliographystyle{mnras}




\appendix

\section{Comparison to Taichi, PH4, and HaloFeedback}

\begin{figure}
    \begin{center}
    \includegraphics[width=0.45\textwidth]{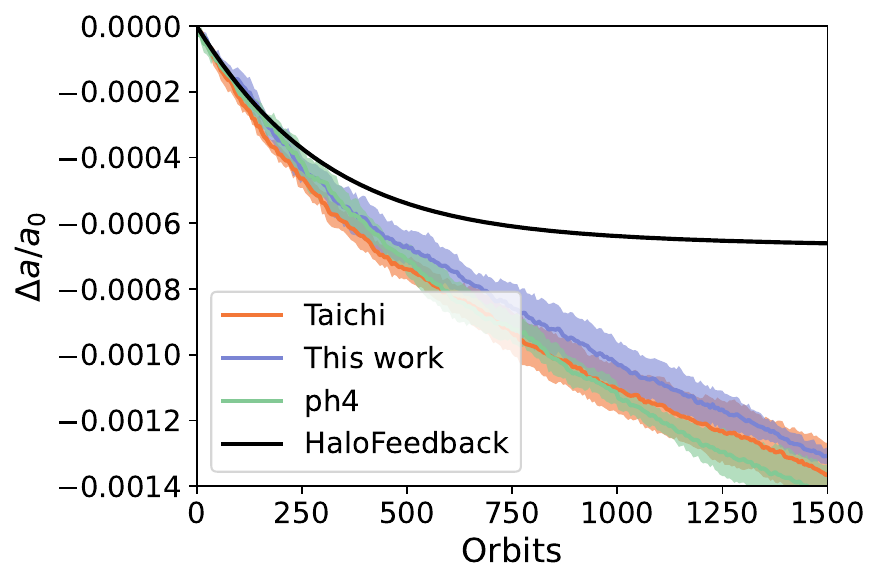}
    \caption{The evolution of the relative change in the semi-major axis $\Delta a /a_0$  as a function of the number of orbits of the binary. The solid lines indicate the average of five independent simulations whereas the shaded region indicates the standard deviation. We find that the results from method presented in this work (purple line) agree with that from {\tt\string Taichi} (orange line), {\tt\string ph4} (green line), $N$-body codes where the self-gravity of the spike is not neglected. This indicates that the effect of the DM-DM interactions is minimal and can be safely ignored, validating our method. Also presented is the evolution of the same system using {\tt\string HaloFeedback} (black line). While the results from {\tt\string HaloFeedback} and the $N$-body codes agree over the first 100-150 orbits, they diverge after that. The spike is able to dissipate more energy from the binary in the $N$-body models compared to the {\tt\string HaloFeedback} models since the impact of DF is subdominant compared to three-body effects.} 
    \label{fig:taichi_compare}
     \end{center}
\end{figure}

To verify the accuracy of our method, we run a set of five simulations with $M_1 = 100 M_{\odot}$, $M_2 = 1 M_{\odot}$, $a_0 = 3 \times 10^{-8}$ pc and $e_0 = 0$. We compare the results to those obtained from $N$-body codes {\tt\string Taichi} \citep[][]{Qirong2021NewA...8501481Z,Mukherjee2021ApJ...916....9M,Mukherjee2023MNRAS.518.4801M}, and {\tt\string ph4} from {\tt\string AMUSE} \citep{Simon2013CoPhC.184..456P,Simon2018araa.book.....P} where the self gravity of the spike is not neglected. {\tt\string Taichi} is run in the direct force summation mode with the fourth order Hamiltonian splitting integrator {\tt\string HHS-FSI} while. The simulations are run for a duration of 1500 orbits of the secondary around the primary. Additionally, we compare our results to those obtained from {\tt\string HaloFeedback} \citep{Kavanagh2020PhRvD.102h3006K} as well since our binary is in a circular orbit initially. We note that we do not include the relativistic terms in the above-mentioned simulations.  

As a test of the validity our method, we present the evolution of the relative change in the semi-major axis $\Delta a / a_0$ from the $N$-body code used in this work, {\tt\string Taichi}, {\tt\string ph4}, and {\tt\string HaloFeedback} in Figure \ref{fig:taichi_compare}. The solid lines represent the mean values obtained from the five simulations whereas the shaded regions represent the standard deviation. We find that the evolution using our approximate force $N$-body method is consistent with that obtained from {\tt\string Taichi} and {\tt\string ph4} where the DM-DM interactions are not neglected. This indicates the effect of the self gravity of the spike is minimal and can be safely neglected. However, we find major differences between the results from the $N$-body codes and {\tt\string HaloFeedback}. The three methods produce consistent results for the first 150 orbits after which the {\tt\string HaloFeedback} results deviate because the binary enters the hard binary phase in which three-body interactions play a dominant role. The binary stalls at $\Delta a/a_0 \approx -5\times 10^{-4}$ in the {\tt\string HaloFeedback models} whereas a longer evolution indicates that the binary actually stalls at $\Delta a/a_0 \gtrsim -2\times 10^{-3}$ in the $N$-body simulations. This is almost $4\times$ larger than what is predicted by {\tt\string HaloFeedback}. As a result of this discrepancy, the dephasing obtained from the {\tt\string HaloFeedback} models should be smaller than those obtained from the $N$-body models, a fact discussed in the results section. The discrepancy arises from the fact that DF is subdominant to three-body scattering as we showed in section 4.3. Moreover, the three-body effects occur over longer timescales, indicating that semi-analytic methods should be validated against long term, secular $N$-body simulations.


\bsp	
\label{lastpage}
\end{document}